\documentclass[number,preprint,3p]{elsarticle}

\usepackage{amssymb}

\usepackage{graphicx}
\usepackage{xcolor}
\usepackage{multicol}
\usepackage{threeparttable}
\usepackage{booktabs}
\usepackage{longtable}
\usepackage{hyperref}
\usepackage{amsthm, amsmath}
\usepackage{mathrsfs}  
\usepackage{algorithm}
\usepackage{algpseudocode}
\usepackage{placeins}
\usepackage{siunitx}
\usepackage{amsfonts}
\usepackage{pgfplots}
\usepackage{subcaption}
\usepackage{enumitem}
\usepackage{setspace}
\newcommand{\beq}{\begin{equation}}
\newcommand{\eeq}{\end{equation}}

\def\ten   #1{\boldsymbol{#1}}
\def\vec   #1{\boldsymbol{#1}}
\usepackage{booktabs}
\usepackage{multirow}
\usepackage{amsmath}
\usepackage{wasysym}

\usepackage{nomencl}
\makenomenclature
\usepackage{etoolbox}
\renewcommand\nomgroup[1]{%
  \item[\bfseries
  \ifstrequal{#1}{A}{Mass, volume, density variables}{%
  \ifstrequal{#1}{B}{Tensors and vectors}{%
  \ifstrequal{#1}{C}{Scalar-valued functions}{}}}%
]}
\usepackage{mdframed}
\usepackage{pgfplotstable}
\usepackage{ifthen}

\journal{arXiv}
\pgfplotsset{compat=1.18}
\begin{document}

\begin{frontmatter}


\title{Collagen and myocyte interplay in cardiac volume overload: a multi-constituent growth and remodeling framework}

\author[inst1,inst2]{Ludovica Maga}
\author[inst1]{Mathias Peirlinck}
\author[inst2]{Lise Noël}

\affiliation[inst1]{organization={Department of BioMechanical Engineering, Delft University of Technology}, addressline={Delft}, country={The Netherlands}
}
\affiliation[inst2]{organization={Department of Precision and Microsystems Engineering, Delft University of Technology}, addressline={Delft}, country={The Netherlands}
}
\begin{abstract}
Hearts subjected to volume overload (VO) are prone to detrimental anatomical and functional chan\-ges in response to elevated mechanical loading, ultimately leading to heart failure. Experimental findings now emphasize that organ-scale changes following VO cannot be explained by myocyte growth alone, as traditionally proposed in the literature. Collagen degradation, in particular, has been associated with VO and assumed
to play a central role in both its acute and chronic stages. 
This hypothesis, however, remains to be substantiated by comprehensive mechanistic evidence, and each constituent contribution to myocardial growth and remodeling (G\&R) processes is yet to be quantified.
In this work, we present a multi-constituent G\&R framework that integrates a mixture-based constitutive model within the kinematic growth formulation. This framework enables us to mechanistically assess the relative contributions of collagen and myocyte changes to alterations in tissue properties, ventricular dimensions, and growth phenotype.
Our numerical results confirm that collagen remodeling affects the passive mechanical response of the myocardium, whereas myocytes predominantly influence the extent and phenotype of VO-induced growth. 
Importantly, collagen degradation exacerbates myocyte hypertrophy, revealing a synergistic interplay that accelerates the left ventricular eccentric growth and thereby promotes systolic dysfunction.
This work constitutes an important step towards an integrated characterization of the early compensatory stages of VO-induced cardiac G\&R.
\end{abstract}


\end{frontmatter}

\section{Introduction}
\label{sec:introduction}
Heart failure remains a leading cause of morbidity and mortality worldwide \citep{Savarese2022}. A frequent pathway to heart failure is volume overload (VO), which refers to an altered mechanical state characterized by retrograde blood flow into the left ventricle (LV). Among various etiologies, VO commonly arises from valvular diseases, such as mitral or aortic regurgitation, which induce higher blood volumes in the ventricle, and thus, elevated wall stretches \citep{Melenovsky2013}. In response to such mechanical stimuli, growth and remodeling (G\&R) processes attempt to compensate for the non-physiological load, i.e., the heart undergoes anatomical and functional changes while preserving its pump function. None\-theless, when persisting over time, VO yields irreversible changes that lead to decompensation and deterioration of the systolic function \citep{Grossman1980, Hutchinson2010}.
Clinically, the transition from compensation to decompensation through G\&R remains difficult to understand and predict \citep{Witzenburg2017}.

Computational modeling offers a powerful framework for investigating G\&R processes mechanistically, by leveraging theoretical frameworks that link mechanical stimuli to long-term tissue adaptation \citep{Goktepe2011, Lee2016}.
Despite significant advances in G\&R theories, a key challenge remains the identification of mechano-regulated G\&R pathways across scales and, consequently, understanding how changes in tissue microstructure translate into whole-organ manifestations. 
Mechanically-driven changes in VO stem from biological processes in the myocardial primary constituents: myocytes, and collagen-rich extracellular matrix (ECM) \citep{Melenovsky2013, Pathak2001}.
Accounting for 70\% to 75\% in cell volume and 25\% to 30\% in cell number of the adult myocardium, myocytes are responsible for volumetric changes in the LV. In response to sustained VO, myocytes deposit sarcomere units in series, leading to lengthening and eccentric hypertrophy \citep{Nakamura2018}. Meanwhile, the ECM provides structural support to the myocardium and serves as scaffolding for myocardial cells, thus regulating mechanical force transmission and defining
passive tissue stiffness \citep{Weber1989}. Extracellular collagen loss alters the constitutive properties of the tissue and increases its compliance \citep{Spinale2000, Cleutjens1996, Hutchinson2010}. Importantly, multiple studies report
that collagen loss occurs early after VO induction (within the first days to weeks) and may precede, and possibly facilitate, progressive myocyte elongation and ventricular dilatation in the compensated phase (first 15 to 20 weeks) \citep{Weber1989, Ryan2007, Hutchinson2010}
These observations indicate that myocyte and collagen-related modifications are both pivotal to the course of G\&R, which thus reflects coupled volumetric and constitutive adaptations.

Yet, few cardiac tissue G\&R modeling approaches intrinsically couple volumetric
growth with evolving constitutive changes \citep{Zhuan2019CoupledInfarction}. 
In this scenario, the vast majority of cardiac VO models rely on the theory of kinematic growth \citep{Kroon2009, Goktepe2011, Genet2016, Peirlinck2019}.  
Within this framework, growth kinematics is fully described by the inelastic component of the deformation gradient \citep{Rodriguez1994}, representing LV volumetric adaptation through the addition of sarcomeres \citep{Peirlinck2019}.
Although these models effectively resolve the LV geometric phenotype, they cannot accurately capture the interplay of G\&R processes 
at the tissue constituent level 
\citep{Lee2016}, nor can they account for evolving properties due to collagen turnover.
On the other hand, constrained mixture (CM) theory offers an alternative to model G\&R in biological tissues and track
each tissue constituent individually. Within CM, each constituent is associated with distinct deposition and removal rates, and the mixture is bound to deform as a single continuum. Despite its more mechanistic nature, CM models pose practical challenges in terms of computational complexity, as each constituent individual evolving reference configuration needs to be tracked over time. Moreover, the framework depends on parameters that are rarely accessible experimentally \citep{Laubrie2022}. 
Reducing the computational cost of CM models motivated the development of homogenized formulations \citep{Humphrey2002}, and remains an active area of research \citep{Gebauer2024b}. Although CM has been largely applied in the context of vascular tissues, its suitability for cardiac tissue has only been scarcely explored to date \citep{Guan2023, Gebauer2023}. 
Importantly, classical CM formulations are often posed without inelastic volumetric growth, such that changes in constituent mass density do not manifest as tissue volume changes \citep{Humphrey2002, Kroon2009AAneurysms}. When volumetric growth is included, it is most often coupled with simplified mechanical membrane models \citep{Baek2006AAneurysms, Figueroa2009ASimulations, Wilson2013ParametricAneurysms}. Only recently, homogenized CM formulations, coupling mass density evolution to volumetric growth, have been proposed. In these approaches, volumetric adaptation is either inferred from the total density increase of all constituents \citep{Braeu2017HomogenizedRemodeling} or introduced through an elastic-swelling, that enforces incompressible growth \citep{Braeu2019AnisotropicTissues}, \citep{Gebauer2023}. 
However, early LV adaptation in response to VO presents a mechano-biological scenario in which volumetric and density-based growth processes are associated with the distinct roles of myocytes and collagen: myocytes predominantly drive volumetric chan\-ges, while collagen primarily governs density modifications in constitutive properties. Accordingly, when modeling VO-driven G\&R, we seek a unified framework that captures both volumetric growth and density adaptation at the individual constituents level.

To address this need, we propose a multi-constituent G\&R computational framework that explicitly distinguishes between myocardial constituents driving volume grow\-th versus density changes.
The framework combines kinematic growth with a mixture-based constitutive formulation, following \citep{Schmid2012}. The key aspect of this approach is the assumption that changes in constituent mass affect tissue behavior through two distinct mechanisms: (i) \emph{constituent-specific volume adaptation} (VA), where added or removed mass changes the constituent volume at approximately constant individual density, and (ii) \emph{constituent-specific density adaptation} (DA), where mass changes manifest primarily as modifications in individual density at approximately constant constituent volume.
By separating VA and DA at the constituent level, the framework is well-suited to disentangle volumetric growth from changes in tissue mechanical properties and to quantify the interplay between myocyte growth and collagen remodeling in response to VO. We leverage this framework to mechanistically explore multi-constituent G\&R, focusing on the early compensatory stage of VO adaptation, where changes in both extracellular collagen and myocytes are observed. The aim is to quantify how myocyte hypertrophy and collagen degradation individually and jointly drive LV dilation, eccentric phenotype, and evolving tissue stiffness.

To this end, Section \ref{sec:mathmodel} outlines the fundamentals of the multi-constituent description of G\&R for a generic soft tissue, providing the mathematical formulation for kinematics, constitutive modeling, and growth. In Sec. \ref{sec:cardmodel}, we develop the cardiac tissue model and specify our modeling choices for myocardial constituents adaptation. 

Section \ref{sec:LVmodel} introduces the idealized ventricular geometry, describes the numerical implementation of the G\&R model, and formalizes the numerical case studies, including three multi-constituent G\&R scenarios: collagen-only (C), myocyte-only (M), and combined (C+M). We present the numerical results in Sec. \ref{sec:numresults}. Section \ref{sec:discussion} discusses the results and provides an outlook for future work.

\section{Multi-constituent tissue mechanics}
\label{sec:mathmodel}
To characterize constituent-specific volume and density adaptation processes in soft tissues, we propose a framework that integrates concepts from finite strain theory, soft tissue constitutive modeling, and G\&R formulations. The subsequent sections detail the relevant governing equations for G\&R, starting with the kinematic and balance equations in Subsecs. \ref{subsec:Kinematics} and \ref{subsec:Balance}. 
Subsection \ref{subsec:Constitutive Model} focuses on the passive material response relying on the definition of the total Helmholtz free energy. 
Based on the theory of mixtures, Subsection \ref{subsec:Multi-constituent formulation of myocardial tissue} extends the framework to a multi-constituent description. Such a description allows for capturing any change in mass, modeled as VA or DA for each constituent, as outlined in Subsec. \ref{subsec: growth and remodeling}.

\subsection{Kinematics}
\label{subsec:Kinematics}
In a three-dimensional domain, the mapping, $\varphi(\vec{X}) = \vec{x}$, maps the location $\vec{X}$ of a point in the reference configuration $\Omega_0$ at time $t = 0$ to its location $\vec{x}$ in the deformed configuration $\Omega_t$ at time $t$. The deformation gradient, $\ten{F} = \partial \vec{x} /\partial \vec{X}$, relates the undeformed and deformed vectors, $d\vec{X}$ and $d\vec{x}$, respectively, and its determinant, $\text{det}\ten{F} = J$, accounts for volume changes. This definition leads to the derivation of the right Cauchy-Green $\ten{C}$ and the left Cauchy-Green tensors $\ten{b}$,
\begin{equation}
    \ten{C} = \ten{F}^\top\ten{F}, \quad 
    \ten{b} = \ten{F}\ten{F}^\top.
\end{equation}

\subsubsection{Kinematics of growth}
\label{subsubsec: Kinematics of growth}
\begin{figure*}[h]
\centering
\includegraphics[width=\textwidth]{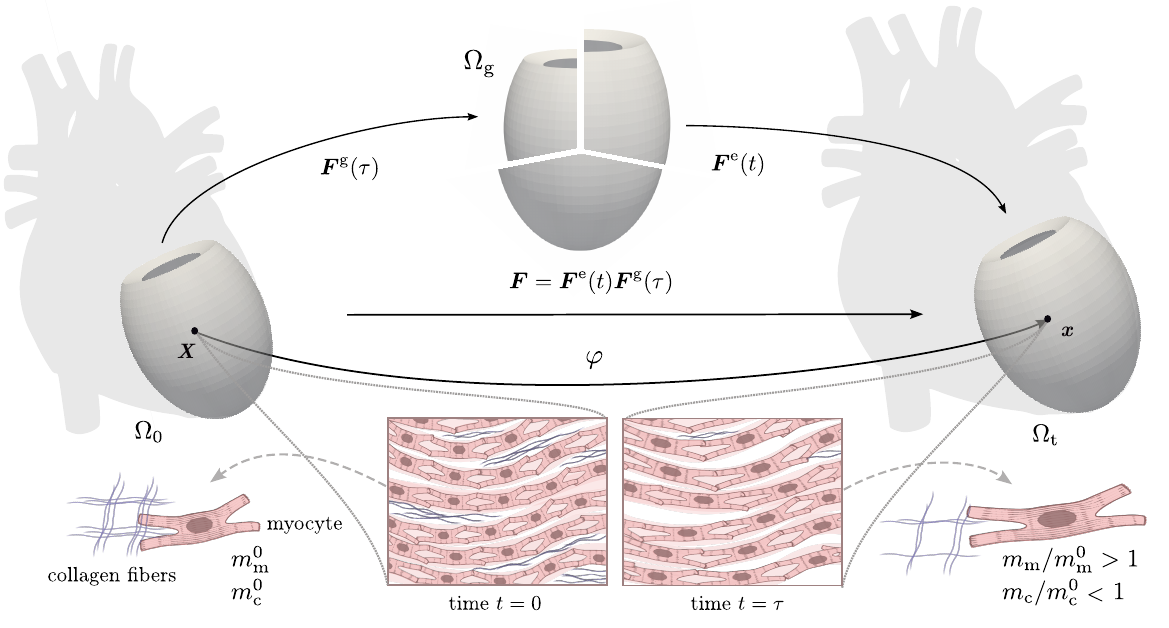}
\caption{\textbf{Kinematics of multi-constituent myocardial tissue G\&R.} A point located at $\vec{X}$ in the reference configuration $\Omega_0$ is mapped to the current configuration $\Omega_t$, such that $\vec{x}=\varphi(\vec{X})$, and undergoes a total deformation characterized by the deformation gradient $\ten{F}$. The total deformation is multiplicatively decomposed into the growth $\ten{F}^\text{g}$ and the elastic $\ten{F}^\text{e}$ part, such that $\ten{F} = \ten{F}^\text{e} \ten{F}^\text{g}$. Growth induces an intermediate, stress-free configuration, $\Omega_\text{g}$, in which the tissue expands locally without mechanical constraints, possibly creating overlaps and incompatibilities. The elastic deformation $\ten{F}^\text{e}$ restores tissue continuity and leads to the residually-stressed configuration $\Omega_t$. At the tissue scale, collagen fibers gradually degrade over time, leading to a reduction in mass. As collagen degrades, interstitial areas appear increasingly devoid of collagen. Myocytes elongate by adding sarcomeres in series, thereby increasing their mass at the end of the G\&R process.}
\label{fig:vo_schematics}
\end{figure*}

To account for the VA, occurring over a longer G\&R time scale $\tau$, and the elastic response, associated with a short time scale $t \ll \tau$, the deformation gradient tensor is decoupled into a inelastic growth part $\ten{F}^\text{g}$ and an elastic part $\ten{F}^\text{e}$ \citep{Rodriguez1994},
\begin{equation}
    \ten{F}=\ten{F}^\text{e}(t) \ten{F}^\text{g}(\tau).
    \label{eq:kinematic_growth}
\end{equation} 
As shown in Fig. \ref{fig:vo_schematics}, G\&R processes change the stress-free reference configuration of the tissue, $\Omega_0$, leading, in general, to a mechanically \textit{incompatible} configuration, $\Omega_g$, introduced through the inelastic growth deformation gradient tensor. 
The geometric and elastic compatibility is restored by the elastic deformation gradient tensor, which generates residual stresses and strains \citep{Goriely2017}. 
Figure \ref{fig:vo_schematics} provides a schematic illustration of the growth kinematics. 
For conciseness, we omit the time dependency of the tensors in the following sections. The elastic right Cauchy-Green tensor $\ten{C}^\text{e}$ is introduced:
\begin{equation}
    \ten{C}^\text{e} = {\ten{F}^\text{e}}^\top \ten{F}^\text{e} = {\ten{F}^\text{g}}^{-\top} \ten{C} {\ten{F}^\text{g}}^{-1}.
\end{equation}
\subsubsection{Kinematics of near-incompressibility}
\label{subsubsec:Kinematics of near incompressibility}
Due to their high water content, biological soft tissues are typically treated as nearly incompressible materials, undergoing minimal elastic volumetric changes in response to external mechanical loading.
We enforce the condition of near-incompres\-sibility through an additional split of the elastic deformation gradient tensor 
into its volumetric $\ten{F}_\text{vol}$ and isochoric part $\Bar{\ten{F}}$: 
\begin{equation}
    \ten{F}^\text{e} = \ten{F}_\text{vol} \Bar{\ten{F}},
    \label{eq: volumetric decomposition}
\end{equation}
where
\begin{equation}
    \ten{F}_\text{vol}= (J^\text{e})^{\frac{1}{3}}\ten{I}, \qquad
    \Bar{\ten{F}} = (J^\text{e})^{-\frac{1}{3}} \ten{F}^\text{e},
\end{equation}
implying that $J=\text{det}\ten{F}$ and $J^\text{e}=\text{det}\ten{F}^\text{e}$, $\text{det}\Bar{\ten{F}} \equiv 1$.
Thus, the isochoric elastic right Cauchy-Green deformation tensor $\Bar{\ten{C}}$ can be further written as: 
\begin{equation}
    \Bar{\ten{C}} = \Bar{\ten{F}}^\top \Bar{\ten{F}} = (J^\text{e})^{-\frac{2}{3}} \ten{C}^{\text{e}},
\end{equation}
such that $\ten{C}^\text{e} = (J^\text{e})^{\frac{2}{3}}\Bar{\ten{C}}$. Accordingly, the isochoric invariants are given as $\Bar{I}_1 = \text{tr}\Bar{\ten{C}}$, and $\Bar{I}_3=\text{det}\Bar{\ten{C}}\equiv 1$. 
The pseudo-invariant is defined as $\Bar{I}_4 = \Bar{\ten{C}}:\ten{N}$, with $\ten{N} = \vec{n} \otimes \vec{n}$, the structural tensor defined by the characteristic material direction $\vec{n}$. 
Tensor product notations are introduced for the dyadic product \{$ \CIRCLE \otimes \Circle \}_{ij} = \{\CIRCLE\}_{i}\{\Circle\}_j$, and the double contraction operation \{$ \CIRCLE:\Circle \} =  \{\CIRCLE\}_{ij}\{\Circle\}_{ij}$.
Near-incompressibility is achieved by introducing a penalty function in the material model, see Subsec. \ref{subsubsec: Constitutive equations of near-incompressibility}.

\subsection{Balance of linear momentum}
\label{subsec:Balance}
Given that G\&R processes typically occur on a time scale of days, weeks, or months, and elastic deformation occurs on a time scale of seconds or less, we can adopt the \textit{slow growth} assumption. The inertia effect from the added mass is therefore negligible, allowing us to consider our system to be quasi-static \citep{Goriely2017}. 
Denoting the body force vector as $\vec{B}_0$, the balance of linear momentum in the reference configuration can be expressed as:
\begin{equation}
    \text{Div} \ten{P} + \vec{B}_0 = \ten{0},
    \label{eq:linear momentum balance}
\end{equation}
with $\ten{P}$ the first Piola-Kirkchhoff stress tensor, and $\text{Div}$ the divergence operator.

\subsection{Constitutive model}
\label{subsec:Constitutive Model}
We adopt a hyperelastic material framework to model biological tissue, with the total deformation represented as the combination of elastic and growth contributions.
Herein, Subsection \ref{subsubsec: Constitutive equations of growth} describes the hyperelastic material model by specifying the total stress and tangent moduli. 
In Sect.~\ref{subsubsec: Constitutive equations of near-incompressibility}, the free energy function is split into a volumetric and an isochoric part, and the constitutive equations are reformulated accordingly. 

\subsubsection{Constitutive equations of growth}
\label{subsubsec: Constitutive equations of growth}
We introduce the Helmholtz free energy function $\psi$, defined per unit volume, and embed the growth kinematics, outlined in Sec. \ref{subsec:Kinematics}. 
It follows that the total second Piola-Kirchhoff stress is defined as:
\begin{equation}
    \ten{S} 
    = 2\frac{\partial{\psi}}{\partial{\ten{C}}} 
    = 2\frac{\partial{\psi}}{\partial{\ten{C}^\text{e}}} : \frac{\partial{\ten{C}^\text{e}}}{\partial{\ten{C}}} 
    = {\ten{F}^\text{g}}^{-1} \ten{S}^\text{e} {\ten{F}^\text{g}}^{-\top}.
    \label{eq: total SPK stress}
\end{equation}
The fourth-order tangent modulus is obtained by differentiating the second Piola-Kirchhoff stress $\ten{S}$ with respect to the right Cauchy-Green tensor $\ten{C}$. 
In a kinematic growth setting, stress is dependent on the total deformation tensor $\ten{F}$ and the growth deformation tensor $\ten{F}^\text{g}$. Hence, the total derivative can be expressed generically through 
\citep{Sez2014}:
\begin{equation}
\begin{split}
    \mathbb{C} 
    &= 2 \frac{d \ten{S}(\ten F^\text{g}, \ten F)}{d \ten{C}} 
    = 2 \frac{\partial \ten{S}}{\partial \ten{C}} \biggr\rvert_{\ten F^\text{g}} 
    + 2 \frac{\partial \ten{S}}{\partial \ten{C}} \biggr\rvert_{\ten{F}} \\
    &= 2 \frac{\partial \ten{S}}{\partial \ten{C}} \biggr\rvert_{\ten F^\text{g}} 
    + 2 \frac{\partial \ten{S}}{\partial \ten F^\text{g}}
    : \frac{\partial \ten F^\text{g}}{\partial \ten{C}}
    \biggr\rvert_{\ten{F}} 
    = \mathbb{C}^\text{e} + \mathbb{C}^\text{g} .
\end{split}
\label{eq:tangent_moduli_growth}
\end{equation}

The first term can be written as:
\begin{equation*}
        \mathbb{C}^\text{e} 
        = (\ten{F}^{\text{g}-1} \overline{\otimes} \ \ten{F}^{\text{g}-1})
         : \mathbb{L}^\text{e}
           : (\ten{F}^{\text{g}-\top} \ \overline{\otimes} \ \ten{F}^{\text{g}-\top}),
    \label{eq:elastic_tangent}
\end{equation*}
where we denote $\mathbb{L}^\text{e}= 2\partial{\ten{S}^\text{e}}/{\partial{\ten{C}^\text{e}}}$ the elastic constitutive modulus. The following notation \{$ \CIRCLE \overline\otimes \Circle \}_{ijkl} = \{\CIRCLE\}_{ik}\{\Circle\}_{jl}$ is introduced for the tensor multiplication.

The second term $\mathbb{C}^\text{g} $ 
is the growth tangent and depends on the specific choice of growth variables, the corresponding definition of the growth tensor, and their evolution equations. The growth tangent modulus specialized for the chosen formulation is provided later in Sec.~\ref{subsec:Multi-constituent formulation of myocardial tissue}.

\subsubsection{Constitutive equations of near-incompressibility}
\label{subsubsec: Constitutive equations of near-incompressibility}
The kinematic decoupling of the elastic deformation gradient tensor in Eq. \eqref{eq: volumetric decomposition} allows for the additive split of the strain energy density function into a volumetric $\psi_\text{vol}$ and an isochoric $\Bar{\psi}$ contribution,
\begin{equation}
    \psi(\ten{C}^\text{e}) = \psi_\text{vol} (J^\text{e}) + \Bar{\psi}(\Bar{\ten{C}}).
    \label{eq: strain energy}
\end{equation}
The first is commonly associated with the presence of water in biological tissue, while the second is defined on the basis of the material microstructure. 
In this work, the volumetric term, chosen to enforce near-incompres\-sibility, is defined as: 
\begin{equation}
    \psi_\text{vol}(J^\text{e}) = \frac{\mu}{2} (J^\text{e} - 1)^2, 
    \label{eq: sedf volumetric}
\end{equation}
where $\mu$ acts as a penalty parameter.

The elastic second Piola-Kirchhoff stress is then defined in terms of its volumetric $\ten{S}_\text{vol}$ and isochoric $\ten{S}_\text{iso}$ parts as: 
\begin{equation}
    \ten{S}^\text{e} 
    = 2\frac{\partial{\psi}(\ten{C}^\text{e})} {\partial\ten{C}^\text{e}} 
    = 2\frac{\partial{\psi}_\text{vol}(J^\text{e})} {\partial\ten{C}^\text{e}} + 2\frac{\partial\Bar{\psi}(\Bar{\ten{C}})} {\partial\ten{C}^\text{e}} 
    = \ten{S}_\text{vol} + \ten{S}_\text{iso},
    \label{eq: elastic stress}
\end{equation}
further defined as:
\begin{equation}
    \ten{S}_\text{vol} 
    =  2\frac{\partial{\psi}_\text{vol}(J^\text{e})}{\partial{J}^\text{e}} : \frac{\partial{J}^\text{e}}{\partial\ten{C}^{e}},
    \label{eq: volumetric stress}
\end{equation}
\begin{equation}
    \ten{S}_\text{iso} 
    = 2\frac{\partial\Bar{\psi}(\Bar{\ten{C}})} {\partial\Bar{\ten{C}}} : \frac{\partial{\Bar{\ten{C}}}} {\partial\ten{C}^\text{e}}
    = \Bar{\ten{S}} : \frac{\partial{\Bar{\ten{C}}}} {\partial\ten{C}^\text{e}},
    \label{eq: isochoric stress}
\end{equation}
where $\Bar{\ten{S}}$ is the fictitious second Piola-Kirchhoff stress.
Analogously, the derivative of the elastic second Piola-Kirchhoff stress with respect to the right-Cauchy strain tensor can be decoupled into a volumetric $\mathbb{L}_\text{vol} $ and isochoric $\mathbb{L}_\text{iso}$ part:
\begin{equation}
    \mathbb{L}^\text{e}  
    = 2 \frac{\partial\ten{S}^\text{e}}{\partial\ten{C}^\text{e}} 
    = 2 \frac{\partial\ten{S}_\text{vol}}{\partial\ten{C}^\text{e}} + 2 \frac{\partial\ten{S}_{\text{iso}}}{\partial\ten{C}^\text{e}} 
    = \mathbb{L}_\text{vol} + \mathbb{L}_\text{iso},
    \label{eq: elastic tangent moduli}
\end{equation}

that can be further specified following \cite{Holzapfel2006}:
\begin{equation}
    \mathbb{L}_\text{vol} 
    = 4 \frac{\partial^2\psi_{\text{vol}}(J^\text{e})}{\partial\ten{C}^\text{e}\partial\ten{C}^\text{e}},
    \label{eq: volumetric elasticity tensor}
\end{equation}
\begin{equation}
    \mathbb{L}_{\text{iso}} 
    = 4\frac{\partial^2\Bar{\psi}(\Bar{\ten{C}})}{\partial\ten{C}^\text{e}\partial\ten{C}^\text{e}}.
    \label{eq: isochoric elasticity tensor}
\end{equation}
The full derivation of the fourth-order elasticity tensor is provided in Appendix 7.

\subsection{Mixture model}
\label{subsec:Multi-constituent formulation of myocardial tissue}
We model soft tissues as a mixture of $I$ constituents at affine deformations, i.e, the constituents
are constrained to follow the deformation of the continuum.
At a material point, all constituents are simultaneously present. 
On a short time scale $t$, the tissue is considered nearly incompressible; however, on a longer time scale $\tau$, it may experience a change in volume due to an increase or decrease in mass. 
The multi-constituent representation allows for changes in individual mass and volume to contribute to the total change of volume of the tissue $V$ \citep{Schmid2012}. 
At the current G\&R time $\tau$, we define the total change in mass and in volume for the tissue as:
\begin{equation}
    \hat{m} = \frac{m}{m^0}, \quad
    \hat{v}= \frac{V}{V^0}.
\end{equation}
For each constituent, we introduce the volume $V_i$, the mass $m_i$, the volume fraction $\phi_i = V_i/V$, and the partial and individual densities, $\rho_i$ and $r_i$. 
At time $\tau = 0$, the latter are defined as:
\begin{equation}
    \rho_{i}^0 = \frac{m_{i}^0}{V^0}, \quad
    r_{i}^0 = \frac{m_{i}^0}{ V_{i}^0}.
\end{equation}
At the current G\&R time $\tau$, we define the normalized mass $\hat{m}_{i}$, the normalized volume $\hat{v}_{i}$, and the partial $\hat{\rho}_i$ and individual $\hat{r}_i$ densities for each constituent with respect to the initial state:
\begin{equation}
    \hat{m}_{i} = \frac{m_{i}}{m_{i}^0}, \quad
    \hat{v}_{i} = \frac{V_{i}}{V_{i}^0}, \quad
    \hat{\rho}_{i} = \frac{\rho_{i}}{\rho_{i}^0}, \quad
    \hat{r}_{i} = \frac{r_{i}}{r_{i}^0}.
\end{equation}

The strain energy per unit volume can be written as the sum of constituent-specific strain energy functions $\Bar{\psi}_i$, weighted by their partial mass density \cite{Humphrey2002}:
    \begin{equation}
    \Bar{\psi}(\ten{\Bar{C}}) = \sum_{i = 1}^{I} \hat{\rho}_i \Bar{\psi}_i. 
    \end{equation}
The isochoric part of the second Piola-Kirchhoff stress reads: 
\begin{equation}
    \ten{S}_\text{iso} 
    = 2\frac{\partial\Bar{\psi}(\Bar{\ten{C}})} {\partial\ten{C}^\text{e}} 
    = 2 \sum_{i = 1}^{I} \hat{\rho}_i\frac{\partial\Bar{\psi}_i(\Bar{\ten{C}})} {\partial\ten{C}^\text{e}}, 
\end{equation}
Following Eq. \eqref{eq: isochoric stress}, it is convenient to define the fictitious second Piola Kirchhoff stress $\Bar{\ten{S}_i}$ associated with each constituent, as:
\begin{equation}
    \Bar{\ten{S}}_i = 2 \hat{\rho_i} \frac{\partial\Bar{\psi}(\Bar{\ten{C}})} {\partial\Bar{\ten{C}}} 
\end{equation}
The fourth-order elasticity tensor is defined as:
\begin{equation}
    \mathbb{L}_\text{iso} 
    = 2\frac{\partial\ten{S}_\text{iso}(\Bar{\ten{C}})}{\partial\ten{C}^\text{e}} 
    = 4 \sum_{i = 1}^{I} \hat{\rho}_i\frac{\partial^2\Bar{\psi}_i(\Bar{\ten{C}})} {\partial\ten{C}^\text{e} \partial\ten{C}^\text{e}}. 
\end{equation}
\subsection{Growth and Remodeling - Volume versus density adaptation}
\label{subsec: growth and remodeling}
We assume that a constituent mass change results in either a change in volume or in density for that constituent \citep{Eriksson2014, Grytsan2017}.  
For VA, the density of the individual constituent is constant while its volume changes. For DA, the density of the individual constituent adapts and its volume remains constant. Mechano-biological considerations substantiate the hypotheses regarding VA and DA for each constituent in Sec. \ref{sec:cardmodel}.

In the VA scenario, the individual mass density remains constant during G\&R, i.e., $r_{i} \equiv r_{i}^0$. 
Therefore, we can write:
\begin{equation}
    \hat{m}_{i} 
    = \frac{m_{i}}{m_{i}^0} 
    = \frac{r_{i} V_{i}}{r_{i}^0 V_{i}^{0}}
    = \frac{V_{i}}{V_{i}^{0}}=\hat{v}_{i}.
\end{equation}
According to \cite{Eriksson2014}, the partial density change can then be written as:
\begin{equation}
    \hat{\rho}_{i}
    = \frac{\rho_{i}}{\rho_{i}^0}
    = \frac{m_{i} V^0}{m_{i}^0 V} 
    = \hat{m}_{i} \frac{1}{\hat{v}} 
    = \hat{v}_{i} \frac{1}{\hat{v}}. 
    \label{eq: partial mass density change}
\end{equation}
In a DA scenario, the constituent-specific mass change is attributed to pure alterations in individual mass density, not affecting the individual volume, i.e., $V_{i} \equiv V_{i}^0$, and it follows that:
\begin{equation}
    \hat{m}_{i}
    = \frac{m_{i}}{m_{i}^0} 
    = \frac{r_{i}V_{i}}{r_{i}^0V_{i}^0} 
    = \frac{r_{i}}{r_{i}^0} 
    = \hat{r}_{i}.
\end{equation}
Similarly to Eq. \eqref{eq: partial mass density change}, the individual mass density change can be written as:
\begin{equation}
    \hat{\rho}_{i} = \hat{r}_{i} \frac{1}{\hat{v}}.
\end{equation}

If some constituents follow VA while others follow DA, the volume change of the whole tissue is given as:
\begin{equation}
    \hat{v}= \frac{V}{V^0}=  \sum_{i=1}^{I} \frac{V_{i}}{V^0}  =\sum_{i=1}^{I} \hat{v}_{i} \phi_{i}^0 
    = \sum_{i_{\text{\tiny VA}}} \hat{m}_{i} \phi_{i}^0 + \sum_{i_{\text{\tiny DA}}}  \phi_{i}^0, 
    \label{eq:volume_change}
\end{equation}
where $\hat{v}_{i} \equiv 1$, if constituent $i$ follows DA, and $\hat{v}_{i} \equiv \hat{m}_{i}$, if it undergoes VA.
The normalized density keeps the generic form:
\begin{equation}
    \hat{\rho}_{i}=\hat{m}_{i} \frac{1}{\hat{v}}.
    \label{eq:partial_density_general}
\end{equation}

G\&R is, thus, modeled such that the change in tissue volume $\hat{v}$ specifies growth kinematics, while the partial density change $\hat{\rho}_{i}$ of an individual constituent $i$ is used to connect the adaptation of the tissue to its elastic response.

Mass growth is typically coupled to a mechanical stimulus through an evolution equation that dictates the rate at which growth occurs over a long time scale. 
The dependency on the mechanical driving force usually involves mechanical stress or strain, which are used to define the homeostatic state, i.e., the preferred state at which the tissue is at equilibrium. 
A generic mass evolution equation is given as: 
\begin{equation}
    \frac{\partial\hat{m}_i}{\partial \tau} \left (\hat{m}_i, {\ten{S^\text{e}}}, \Bar{\ten{C}}\right ) = \kappa_i\, \xi_i (\hat{m}_i, {\ten{S^\text{e}}}, \Bar{\ten{C}}),
    \label{eq: general_stimulus}
\end{equation}
where the growth criterion function $\xi_i$ for a constituent $i$ is defined to promote the restoration to the homeostatic state. 
The temporal evolution is determined by the growth gain constant $\kappa_i$.\

The definition of the tangent modulus introduced in Sect.~\ref{subsubsec: Constitutive equations of growth} is further specified for the presented G\&R formulation. 
In our multi-constituent framework, G\&R is described through constituent-specific mass changes. 
We, therefore, select the deformation tensor $\ten{F}$ and the individual mass ratios $\hat{m}_i$ as the main dependencies. 
For a fully coupled linearization of the G\&R problem, the corresponding growth material tangent reads:
\begin{equation}
\mathbb{C}^\text{g} = 2 \frac{\partial \ten{S}}{\partial \ten F^\text{g}}  :  \sum_{i=1}^{I} \frac{\partial \ten F^\text{g}}{\partial \hat m_i} \otimes \frac{\partial \hat m_i}{\partial \ten{C}} \biggr\rvert_{\ten{F}}.
\label{eq: growth component elasticity tensor multi-const}
\end{equation}
In the numerical scheme employed in this work, however, $\mathbb{C}^\text{g}$ is not included in the global displacement linearization, as detailed in Sect.~\ref{subsec:numerical}. 
\section{Myocardial tissue model} 
\label{sec:cardmodel}
In the following subsections, the generic constitutive framework introduced in Subsec. \ref{subsec:Constitutive Model} is specialized for myocardial tissue. We briefly summarize the key morphological characteristics of the myocardium and its principal structural constituents. 
In Subsec. \ref{subsubsec:Myocardial tissue mixture constitutive model}, we define the strain energy density functions associated with each constituent. The choice of the initial set of constitutive parameter used in the model is specified in Subsec. \ref{subsubsec:Parameter Identification}.
Subsection \ref{subsubsec:constituent_mass_change} outlines the evolution equations governing mass growth for each constituent, and the mechanically-driven stimuli acting as the G\&R driving factors are specified.
The growth deformation gradient tensor is introduced in Subsec. \ref{subsubsec:Myocardial tissue growth}, and is formulated to capture the eccentric remodeling pattern observed at the organ level under VO conditions. 

\subsection{Myocardial mixture-based constitutive model}
\label{subsubsec:Myocardial tissue mixture constitutive model}
Histologically, the LV myocardium is a layered composite structure made of myocytes and a network of collagen fibers embedded in a ground matrix \citep{Rohmer2007, Xi2019, Wilson2022}. 
This organized architecture in layers, or \textit{sheets}, provides a natural reference to define an orthonormal set of material directions. 
The local basis vectors include the fiber direction, $\vec{f}_0$, which aligns with the longitudinal axis of myocytes, the sheet direction, $\vec{s}_0$, which lies within the transmural plane, and the sheet-normal direction, $\vec{n}_0$, which is orthogonal to both $\vec{f}_0$ and $\vec{s}_0$. Within the collagen network, different fiber families, often referred to as \textit{endomysial} and \textit{perimysial} collagen, form lateral connections both between adjacent myocytes and between sheets.

Following the well-established microstructure-inspired Holzapfel-Ogden constitutive model for myocardial tissue \cite{Holzapfel2009}, we proposed a mixture-based constitutive model defined by additive contributions of each of the myocardial constituents to the tissue's mixture strain energy function:
\begin{equation}
    \Bar{\psi}(\ten{\Bar{C}}) 
    = \sum_{i = 1}^{I} \hat{\rho}_i \Bar{\psi}_i 
    = \hat{\rho}_\text{g} \Bar{\psi}_\text{g} +\hat{\rho}_\text{m} \Bar{\psi}_\text{m} + \hat{\rho}_\text{c} \Bar{\psi}_\text{c}, 
    \label{eq: strain energy constituent}
\end{equation}
where we distinguish between myocytes, $i=m$, collagen network, $i = c$, and ground matrix, $i=g$. 

\textbf{Myocytes}. Myocytes are assumed to bear loads along their principal fiber direction \citep{Liu2025, Gebauer2023}, and therefore, the associated strain energy contribution consists of a one-dimensional exponential term:
\begin{equation}
\Bar{\psi}_\text{m} =\frac{a_\text{m}}{2b_\text{m}} \left(\exp \left(b_\text{m}\langle \Bar{I}_\text{4m}-1\rangle^2 \right)-1\right),
\label{eq: myocytes sedf}
\end{equation}
$a_\text{m}$ and $b_\text{m}$ are material parameters, $\Bar{I}_\text{4m} = \vec{f}_0 \cdot \ten{\Bar{C}} \vec{f}_0$ is the isochoric fourth pseudo-invariant, and $\vec{f}_0$ denotes the preferred direction of myocytes in the reference configuration. The Macaulay brackets are defined such that:
\begin{equation}
\langle \CIRCLE - \Circle \rangle = \left\{ \begin{array}{ll}
\CIRCLE - \Circle, &\quad \text{if } \CIRCLE > \Circle,\\
0, &\quad \text{otherwise.}
\end{array}\right.
\end{equation}

\textbf{Collagen}. Collagen forms a fibrous network that envelops the myocytes. 
Image-based characterization of the architecture of myocardial collagen \citep{Rohmer2007,Wang2016Image-drivenFibrosis} suggests a distinction between endomysial and perimysial collagen fibers. 
Endomysial collagen packs the myofibers together in sheet-like clusters, leading to load-bearing contribution along both the principal fiber direction, $\ten{f}_0$, and the the sheet direction, $\vec{s}_0$.
Moreover, this packing further provides a shear stiffening contribution in the fiber-sheet plane \citep{Sommer2015,Liu2025}.
Perimysial collagen fibers connect the different sheets together, and act along the sheet-normal direction, $\vec{n}_0$. 
These observations motivate the following four-term strain energy function contributions of the collagen fiber network: 
\begin{equation}
\begin{aligned}
 \Bar{\psi}_\text{c} &= 
    \ \frac{a_\text{cf}}{2b_\text{cf}}
        \left(\exp\!\left(b_\text{cf}\langle \Bar{I}_\text{4cf} - 1\rangle^2\right) - 1\right) \\
    &+ \frac{a_\text{cs}}{2b_\text{cs}}
        \left(\exp\!\left(b_\text{cs}\langle\Bar{I}_\text{4cs} - 1\rangle^2\right) - 1\right) \\
    &+ \frac{a_\text{cn}}{2b_\text{cn}}
        \left(\exp\!\left(b_\text{cn}\langle\Bar{I}_\text{4cn} - 1\rangle^2\right) - 1\right) \\
    &+ \frac{a_\text{cfs}}{2b_\text{cfs}}
        \left(\exp\!\left(b_\text{cfs}(\Bar{I}_\text{8cfs})^2\right) - 1\right),
\end{aligned}
\label{eq:collagen_constitutive}
\end{equation}
where $a_\text{cf},$ $a_\text{cs},$ $a_\text{cn},$ $a_\text{cfs}$, $b_\text{cf},$ $b_\text{cs},$ $b_\text{cn},$ and $b_\text{cfs}$ are material parameters. The isochoric pseudo-invariants are defined as $\Bar{I}_\text{4cf} = \vec{f}_0 \cdot \ten{\Bar{C}} \vec{f}_0$, $\Bar{I}_\text{4cs}=\vec{s}_0 \cdot \ten{\Bar{C}} \vec{s}_0$, and $\Bar{I}_\text{4cn}=\vec{n}_0 \cdot \ten{\Bar{C}} \vec{n}_0$. 
The fourth term includes the coupling invariant $\Bar{I}_\text{8cfs} = \vec{f}_0 \cdot \ten{\Bar{C}} \vec{s}_0$.

\textbf{Ground matrix}. Remaining constituents, primarily elastin, are uniformly distributed within the myo\-car\-dium \citep{Liu2025, Xi2019, McEvoy2018} and are often assumed to bring an isotropic contribution to the tissue response \citep{Guan2019OnLaw}. 
We set the ground matrix strain energy contribution as follows:
\begin{equation}
        \Bar{\psi}_\text{g} = \frac{a_\text{g}}{2b_\text{g}} \left(\exp \left(b_\text{g}(\Bar{I}_{1}-3) \right) -1 \right),
    \label{eq: ground matrix}
\end{equation}
where $a_\text{g}$ and $b_\text{g}$ are material parameters.
The stress and elasticity tensor associated with this myocardial mixture model are derived in \hyperref[sec:appendix_d]{Appendix D}, where we also discuss the implications of formulating the anisotropic constituent contributions in terms of modified isochoric invariants.

\subsection{Constitutive parameter identification}
\label{subsubsec:Parameter Identification}
Table \ref{tab:constitutive_parameters} lists the constitutive parameters used for our mixture-based constitutive law defined in Subsec. \ref{subsubsec:Myocardial tissue mixture constitutive model}. 
We work in two stages to calibrate realistic myocardial material properties. First, we use ex vivo triaxial shear test and biaxial tensile test data \citep{Sommer2015} to calibrate our material properties \citep{Martonova2024}, as discussed in \hyperref[sec:appendix_a]{Appendix A}. 
Next, we rescale these parameters to represent realistic in vivo compliance behavior in Subsec. \ref{sec:organParameters}
\begin{table}[h]
\caption{\textbf{Material parameters.} Constitutive parameters, initially optimized for ex vivo experimental data \citep{Sommer2015}, and scaled to match in vivo pressure--volume response described by the Klotz curve \citep{Klotz2006DownloadedLib}. Details on the ex-to-in vivo scaling procedure \citep{Peirlinck2018b} are provided in \hyperref[sec:appendix_a]{Appendix A}.}
\label{tab:constitutive_parameters}
\centering
\small 
\renewcommand{\arraystretch}{1.15} 
\setlength{\tabcolsep}{3pt} 
\begin{tabular}{lccc}
\toprule
\textbf{Parameter} & \textbf{Ex vivo} & \textbf{In vivo (scaled)} & \textbf{Unit} \\
\midrule
\textbf{Myocytes} &  &  &  \\
$a_\text{m}$ & 0.99 & 0.27 & kPa \\
$b_\text{m}$ & 23.7 & 9.32 & -- \\
\cmidrule(lr){1-4}
\textbf{Collagen} &  &  &  \\
$a_\text{cf}$ & 2.32 & 0.67 & kPa \\
$b_\text{cf}$ & 23.70 & 9.32 & -- \\
$a_\text{cs}$ & 1.41 & 0.41 & kPa \\
$b_\text{cs}$ & 20.07 & 7.89 & -- \\
$a_\text{cn}$ & 2.04 & 0.59 & kPa \\
$b_\text{cn}$ & 16.98 & 6.68 & -- \\
$a_\text{cfs}$ & 0.56 & 0.17 & kPa \\
$b_\text{cfs}$ & 1.08 & 0.43 & -- \\
\cmidrule(lr){1-4}
\textbf{Ground matrix} &  &  &  \\
$a_\text{g}$ & 0.95 & 0.27 & kPa \\
$b_\text{g}$ & 5.45 & 2.15 & -- \\
\bottomrule
\end{tabular}
\end{table}

\subsection{Constituent mass change}
\label{subsubsec:constituent_mass_change}
The VA and DA scenarios introduced in Subsec. \ref{subsec: growth and remodeling}  are now specialized for each myocardial constituent. 

\textbf{Myocytes}. Myocytes primarily contribute to cardiac G\&R through \textit{hypertrophy}, wherein the cells increase in size, rather than through \textit{hyperplasia}, which involves an increase in the number of cells \citep{Zhong2006HypertrophicPathway}.
This physiological behavior supports the choice of a VA assumption to represent myocyte growth, see Subsec. \ref{subsec: growth and remodeling}.
Following prior work identifying myofiber stretch as a potential biomechanical stimuli for eccentric hypertrophy \citep{Peirlinck2019}, we model myocyte mass changes with the following stretch-driven evolution law:
\begin{equation}
    \frac{\partial\hat{m}_\text{m}}{\partial \tau} 
    = \frac{\partial\hat{v}_\text{m}}{\partial \tau} 
    = \kappa_\text{m}\, \hat{m}_\text{m}  \langle \Bar{\lambda}_\text{m} - \Bar{\lambda}_{\text{m,hom}} \rangle,
    \label{eq: stimulus myocytes}
\end{equation}
where growth is activated only if the elastic isochoric stretch, $\bar{\lambda}_{\mathrm{m}}
= \sqrt{\vec{f}_0 \cdot \Bar{\ten{F}}^{\top}\Bar{\ten{F}} \vec{f}_0}$, 
exceeds a critical homeostatic stretch level $ \Bar{\lambda}_{\text{m,hom}}$. 
Here, $\kappa_\text{m}$ is a growth rate parameter that regulates the dynamics of myocyte growth.
 
{\textbf{Collagen}. 
In the myocardium, collagen turnover involves a complex cascade of mechano-biological subprocesses, including changes in ECM proteins and their regulation by cellular and mechanical cues, among other factors.
Since the present study focuses on early VO-induced remodeling, where collagen undergoes a net reduction in content, we do not explicitly model the continuous turnover of collagen production and degradation. 
Instead, we prescribe the resulting net collagen mass change, assuming it follows DA \citep{Sesa2023, Saez2013}. 
Collagen remodeling is therefore not mechanically driven in the present formulation but is prescribed through an exponential decay calibrated to the experimentally reported reduction in collagen content under VO conditions  \citep{Ryan2007, Corporan2018}.
We write the collagen mass evolution as: 
 \begin{equation}
    \hat{m}_\text{c}(\tau) 
     = \hat{m}_{\text{c,min}} + \left(\hat{m}_\text{c}^0 - {\hat{m}_{\text{c,min}}} \right) \text{exp}\left(-\tau/\tau_{\text{end}}\right),
    \label{eq:coll_degrad}
\end{equation}
with the growth parameter $\tau_\text{end}$ denoting the G\&R time $\tau$ for the degradation of collagen to reach a minimal mass ratio $\hat{m}_{\text{c,min}}$. 

It should be noted that this simplification holds here, as we focus our analysis on the compensated stage of VO, limited to the 15 to 20 weeks of G\&R \citep{Hutchinson2010} and associated with a net decrease in collagen mass. 
Net collagen mass increase, instead, might occur in later stages, often referred to as decompensated, as an attempt to counterbalance the increased myocardial compliance \citep{Spinale2000}.

\subsection{Volumetric tissue growth}
\label{subsubsec:Myocardial tissue growth}
Any tissue inelastic volumetric change is accommodated by the growth part of the deformation gradient $\ten{F}^\text{g}$. We can write a generic growth deformation gradient of the form:
\begin{equation}
    \ten{F}^\text{g} = \alpha\ten{I} + [\beta - \alpha] \vec{f}_0 \otimes \vec{f}_0,
\end{equation}
where $\alpha$ and $\beta$ are coefficients specifying the amount of volume change along the constituent's main longitudinal and transverse directions, respectively. 
Their values define the growth kinematics and must ensure that $\text{det}\ten{F}^\text{g} \equiv \hat{v}$.
In the case of cardiac VO, it has been shown that myocytes, undergoing VA, mainly add sarcomeres in series, resulting in longitudinal growth that occurs parallel to the main fiber direction\citep{SahliCostabal2019MultiscaleFailure, Peirlinck2019}. 
As such, we set $\alpha=1$ and $\beta=\hat{v}$, leading to: 
\begin{equation}
    \ten{F}^\text{g} = \ten{I} + [\hat{v} - 1] \vec{f}_0\otimes \vec{f}_0.
    \label{eq: transv_growth}
\end{equation}

\section{Cardiac finite element model} 
\label{sec:LVmodel}
In this section, we present the implementation of the proposed G\&R framework in an idealized human LV. 
Section \ref{subsec:left_vetricle} describes the idealized LV geometry, its finite element discretization, and the assignment of myocardial fiber architecture. Section ~\ref{subsec:numerical} contains the numerical treatment of the primary variables and the solution strategy adopted. 
In Sect.~\ref{sec:organParameters}, we rescale the passive material parameters introduced in Sect.~\ref{subsubsec:Parameter Identification} to ensure physiologically realistic baseline LV mechanics at the organ level. 
Finally, we outline the numerical protocol used to simulate VO in Sect.~\ref{subsec:volume_overload}, including the setup of the homeostatic reference state, the induction of pathological loading, and the resulting G\&R response.

\subsection{Idealized left ventricle model}
\label{subsec:left_vetricle}
We implement our multi-constituent G\&R framework in an idealized LV model.  
Figure \ref{fig:geometry_domain_boundaries} shows the LV in its reference configuration, which we approximate through its end-systolic geometry \citep{Peirlinck2018a}.

The idealized LV domain is represented by a thick-walled, prolated ellipsoid with variable thickness, as shown on the left side of the figure. 
The geometry parameters are provided in Table \ref{tab:parameters_geometry}.
We distinguish boundaries associated with the endocardium (inner wall) $\Gamma_{0}^{\text{endo}}$, the epicardium (outer wall) $\Gamma_{0}^{\text{epi}}$, and the base, $\Gamma_{0}^{\text{base}}$. 
On the right of the figure, the local fiber orientation, i.e., the helix angle, with respect to the tangent plane varies transmurally from -$41^\circ$ at the epicardium to $66^\circ$ at the endocardium, as measured through diffusion-tensor MRI on multiple healthy human heart samples \citep{Lombaert2012,Peirlinck2018b}. 
\begin{figure*}[h]\centering
\includegraphics[width=\textwidth]{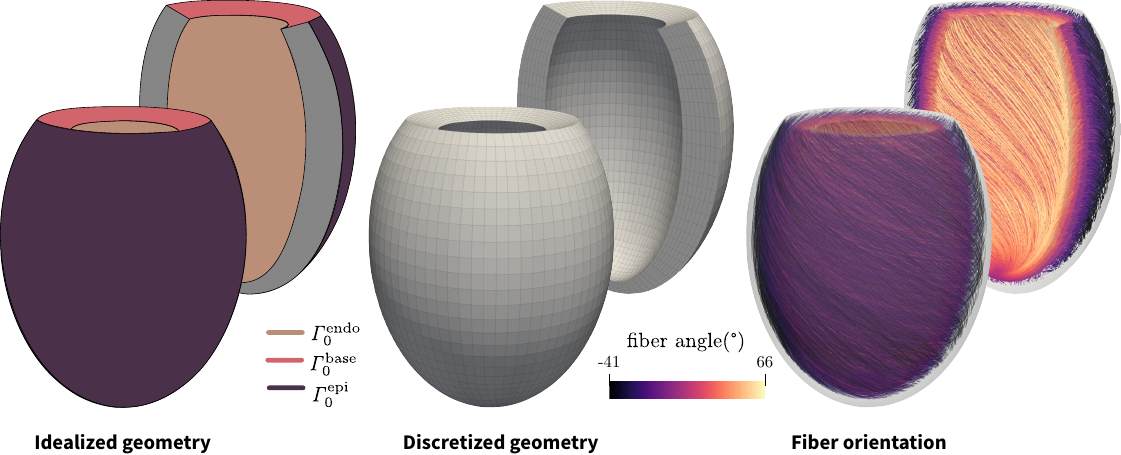}
\caption{\textbf{Idealized left ventricle model}. The LV geometry is approximately represented by an ellipsoid (left). Based on the geometric parametrization, the surface boundaries are identified: the endocardium $\Gamma_{0}^\text{endo}$, the epicardium $\Gamma_{0}^\text{epi}$, and the basal surface $\Gamma_{0}^\text{base}$. The idealized LV domain is discretized using hexahedral finite elements (middle). Lastly, the fiber orientation is retrieved by analytically computing the fiber direction $\vec{f}_0$ (right) forming an angle, the helix angle, with respect to the tangential plane. The helix angle varies transmurally from -$41^\circ$ at the epicardium to $66^\circ$ at the endocardium.} 
\label{fig:geometry_domain_boundaries}
\end{figure*}

The mesh is generated using the open-source geometrical modeling and finite element analysis software \textit{pyFormex} \citep{Peirlinck2017}.
A ventricular mesh consisting of 15,264 first-order Lagrange hexahedral elements was found to be a good trade-off between computational time and accuracy.
Integration of the governing equations is performed with a $3{\times}3{\times}3$ Gauss quadrature rule.
The resulting system of equations is solved with a Newton-Raphson solver using a direct LU factorization available through the Scipy \textit{spsolve} routine, which internally relies on the SuperLU solver \citep{Li2003}.
Governing equations are solved using an in-house software.  
\renewcommand{\arraystretch}{1.15} 
\begin{table}[h]
\caption{\textbf{Left ventricle geometry parameters.}}\label{tab:parameters_geometry}%
\centering
\begin{tabular}{
  l  
  l  
  >{\centering\arraybackslash}m{0.75cm}  
  >{\centering\arraybackslash}m{0.75cm}  
}
\toprule
\textbf{Parameter} & & \textbf{Value} & \textbf{Unit} \\
\midrule
   Major endocardial radius & $r^{\text{endo,l}}$ & 42.9 & mm \\
   Minor endocardial radius & $r^{\text{endo,s}}$ & 20.6 & mm \\
   Apex wall thickness & $t^{\text{apex}}$ & 4 & mm \\
   Mid wall thickness & $t^{\text{mid}}$ & 12 & mm \\
   Epicardial fiber helix angle & $\varphi^{\text{epi}}$ & -41 & deg \\
   Endocardial fiber helix angle & $\varphi^{\text{endo}}$ & 66 & deg\\
\bottomrule
\end{tabular}
\end{table}

\subsection{Numerical implementation} 
\label{subsec:numerical}

In the adopted formulation, the displacement field $\vec{u}$ constitutes the primary unknown, whereas the constituent mass ratios $\hat{m}_i$ are history-dependent internal quantities stored locally at the integration points.}
The displacement field is discretized using standard finite elements, with the corresponding displacement degrees of freedom defined at the nodes. 
Consequently, the system of equations is formulated solely in terms of displacements.

The growth and mechanical responses are coupled at the constitutive level through the mixture-based strain energy function: the growth tensor $\ten{F}^{\mathrm g}$ depends on the mixture volume ratio $\hat{v}$, and, in turn, on the constituent mass ratios $\hat m_i$. The evolution of the latter is driven by an elastic stimulus, directly dependent on the tissue's deformation state through the right Cauchy-Green tensor $\ten C$.

The adopted solution strategy is summarized in Algorithm~\ref{algo:solution}.
The time is discretized into time steps $\Delta t$, such that a time $t_{n+1}$ is given as $t_{n+1} = t_n + \Delta t.$
Within each time step, an updated configuration is iteratively solved for via a Newton-Raphson scheme. 
At each iteration $k$, a trial elastic state is evaluated from the current total deformation and a trial growth tensor 

set to its previous state, such that:
\begin{equation}
\ten F_{n+1}^{\mathrm{g,trial}} = 
\left\{
\begin{array}{ll}
\ten F_{n}^{\mathrm{g}}, & \text{if } k=0,\\
\ten F_{n+1}^{\mathrm{g},(k-1)}, & \text{if } k>0.
\end{array}
\right.
\label{eq:Fgtrial}
\end{equation}
The corresponding trial elastic state, obtained through Eq.~\eqref{eq:kinematic_growth}, determines the current growth stimulus $\xi_i^{(k)}$ from Eq.~\eqref{eq: general_stimulus}. 
The constituent mass ratios are then updated locally through an explicit forward Euler approximation
\begin{equation}
    \hat{m}_{i,n+1}^{(k)}
    =
    \hat{m}_{i,n}
    +
    \Delta t\,\kappa_i\hat{m}_{i,n}\xi_i^{(k)}.
    \label{eq:explicit_mass_update}
\end{equation}
Since Eq.~\eqref{eq:explicit_mass_update} provides $\hat{m}_{i,n+1}^{(k)}$ directly, no local solve is required for the growth internal variables. 
At each iteration $k$, mass ratios are re-evaluated from the previous final value $\hat{m}_{i,n}$ at time $t_{n}$. 
The resulting mass ratios determine the updated mixture volume ratios, growth tensor, and constituent densities, from which the current elastic state and stress are computed.
Growth quantities are held fixed during the subsequent mechanical correction. 
The balance of linear momentum in Eq.~\eqref{eq:linear momentum balance} is written in residual form as $R_u$. Accordingly, the tangent stiffness matrix follows from linearization:
\begin{equation}
    \ten K_{u}^{(k)}
    =
    \left.
    \frac{\partial R_u(\ten u_{n+1}^{(k)})}{\partial\ten u_{n+1}^{(k)}}
    \right|_{
        \hat{m}_{i,n+1}^{(k)}}
    \label{eq:Kutangent}
\end{equation}
and the displacement is updated according to:
\begin{equation}
\ten u_{n+1}^{(k+1)}
    =
    \ten u_{n+1}^{(k)}
    +
    \Delta\ten u^{(k)}
\end{equation}
with the displacement increment:
\begin{equation}
    \Delta\ten u^{(k)}
    =
    -
    \left[
        \ten K_u^{(k)}
    \right]^{-1}
    R_u\left(\ten u_{n+1}^{(k)}\right).
\label{eq:Duupdate}
\end{equation}
The growth update and mechanical correction are repeated until convergence, after which the final iteration mass ratios are carried forward to the subsequent growth time step.
To dictate the convergence of the mechanical problem, a tolerance of $\epsilon = 10^{-8}$ is applied to the displacement-based stopping criterion.

Comparable explicit local growth variable updates have been used in other soft tissue G\&R frameworks~\citep{Klepach2012,Gebauer2023}, provided that a sufficiently small growth time step is selected. 
Numerical robustness with respect to the time step size was assessed by repeating the simulations using increasingly smaller growth time steps, until the resulting growth quantities converged.

Alternative formulations, in which the constituent mass ratios are treated as additional unknowns, have also been proposed~\citep{Kuhl2004}.

\begin{algorithm}[H]
\caption{\footnotesize{Solution strategy for G\&R model}}
\label{alg:gr_scheme_k}
\footnotesize
\setstretch{1.35}
\begin{algorithmic}[1]

\State Given tolerance $\epsilon$, final time $T$, time step $\Delta t$
\State Set $n \gets 0$ and $t_0=0$
\State Initialize $\ten u_{0} \gets \mathbf{0}$, $\hat{m}_{i,0} \gets 1$
\While{$t_n<T$}
    \State $t_{n+1} = t_n+\Delta t$
    \State Set $k \gets 0$
    \State Initialize $\ten u_{n+1}^{(0)} \gets \ten u_n$

    \Repeat
        \State Compute $\ten F_{n+1}^{(k)} = \ten I + \nabla\ten u_{n+1}^{(k)}$
        
        \State Set $\ten F_{n+1}^{\mathrm{g,trial}}$ following Eq.~\eqref{eq:Fgtrial}
        \State Estimate $\ten F_{n+1}^{\mathrm{e,trial}} = 
        \ten F_{n+1}^{(k)}
        \left[
            \ten F_{n+1}^{\mathrm{g,trial}}
        \right]^{-1}$ 
        \State \phantom{Estimate} $\xi_i^{(k)}$ from Eq.~\eqref{eq: general_stimulus}

        \State Update $\hat{m}_{i,n+1}^{(k)}$ from Eq.~\eqref{eq:explicit_mass_update}
        \State \phantom{Update} $\hat{v}_{n+1}^{(k)}$ from Eq.~\eqref{eq:volume_change}
        \State \phantom{Update} $\ten F_{n+1}^{\mathrm{g},(k)}$ from Eq.~\eqref{eq: transv_growth}
        \State \phantom{Update} $\ten F_{n+1}^{\mathrm{e},(k)} = 
        \ten F_{n+1}^{(k)}
        \left[
            \ten F_{n+1}^{\mathrm{g},(k)}
        \right]^{-1}$
        \State \phantom{Update} $\hat{\rho}_{i,n+1}^{(k)}$ from Eq.~\eqref{eq:partial_density_general}
        \State \phantom{Update} $\ten S_{n+1}^{(k)}$ from Eq.~\eqref{eq: total SPK stress}
        
        \State Compute $R_u^{(k)}$ from Eq.~\eqref{eq:linear momentum balance}
        \State \phantom{Compute} $\ten K_u^{(k)}$ from Eq.~\eqref{eq:Kutangent}

        \State \phantom{Compute} $\Delta\ten u^{(k)}$ from Eq.~\eqref{eq:Duupdate}

        \State Update $\ten u_{n+1}^{(k+1)}
        =
        \ten u_{n+1}^{(k)}
        +
        \Delta\ten u^{(k)}$

        \State Set $k \gets k+1$
    \Until{${\lVert \vec{u}_{n+1}^{(k)} - \vec{u}_{n+1}^{(k-1)} \rVert} / {\lVert \vec{u}_{n+1}^{(0)} \rVert}
    <\epsilon,$}

    \State Set $n \gets n+1$
\EndWhile

\end{algorithmic}
\label{algo:solution}
\end{algorithm}

\subsection{In vivo parameter scaling}
\label{sec:organParameters}
Prior work has shown that pure ex vivo calibration of myocardial tissue often leads to overly stiff in vivo passive tissue behavior \citep{Wang2013,Sack2018ConstructionDT-MRI,Peirlinck2018b}. 
To ensure that our baseline LV model reproduces physiologically realistic in vivo tissue compliance, we follow the two-stage ex-to-in vivo tissue calibration protocol in \cite{Peirlinck2018b}, described in more detail in \hyperref[sec:appendix_b]{Appendix B}.

\subsection{Volume overload simulations}
\label{subsec:volume_overload}
To model multi-constituent adaptations throughout the first 15 to 20 weeks of VO-induced G\&R, we work in three sequential stages: 
(i) the pre-loading of the LV to the (baseline) end-diastolic state, 
(ii) the induction of pathological VO by increasing end-diastolic pressure (EDP), and 
(iii) the G\&R progression at the tissue- and the organ-level driven by constituent-specific adaptation laws.
In stage (i), we compute the healthy end-diastolic configuration that serves as the pre-growth homeostatic baseline state. 
Starting from the unloaded reference geometry, the ventricle is inflated quasi-statically by applying a linear ramp of internal pressure from 0\,mmHg to a physiological EDP of 8\,mmHg \citep{Wang2013} on the endocardial surface $\Gamma_{0}^\text{endo}$. 
Zero displacements are prescribed on the basal surface $\Gamma_{0}^\text{base}$, while the epicardial surface $\Gamma_{0}^\text{epi}$ is traction-free.
The resulting end-diastolic isochoric myofiber stretch $\overline{\lambda}_\text{m}$ is stored at each Gauss point and defines the regional homeostatic stretch $\overline{\lambda}_\text{m,hom}$ in Eq.~\eqref{eq: stimulus myocytes}. 
In stage (ii), we impose VO by increasing the EDP to 16\,mmHg \citep{Kerckhoffs2012}, thereby elevating the diastolic wall stretch relative to the homeostatic baseline. 
This overloaded configuration provides the mechanical state from which G\&R is initiated.
In stage (iii), we simulate G\&R over $T=19$~weeks with a time step $\Delta t=0.5$~days.  
During this period, constituent-specific G\&R processes are activated, 
and the evolution of both local passive tissue behavior, e.g., compliance via changes in collagen content, and global organ-scale measures, e.g., chamber dilation and geometry, is tracked. 
Specifically, collagen degradation is modeled by prescribing a decay of collagen mass, as introduced in Eq.~\eqref{eq:coll_degrad}.
G\&R parameter values are listed in Table \ref{tab:parameters_VO}.
\begin{table}[h]
\caption{\textbf{Volume overload simulations parameters.}}\label{tab:parameters_VO}%
\centering
\renewcommand{\arraystretch}{1.15} 
\begin{tabular}{
  l  
  l  
  >{\centering\arraybackslash}m{0.75cm}  
  >{\centering\arraybackslash}m{0.75cm}  
}
\toprule
\textbf{Parameter} & & \textbf{Value} & \textbf{Unit} \\
\midrule
\textbf{Boundary conditions} &&&\\ 
\midrule
  Preload diastolic pressure & $P_{\text{pre}}$ & 8 & mmHg \\
  Overload diastolic pressure & $P_{\text{over}}$ & 16 & mmHg \\[5pt]
\textbf{Initial volume fractions} &&&\\ 
\midrule
  Myocyte  & $\phi^0_\text{m}$ & 0.700 & -- \\
  Collagen  & $\phi^0_\text{c}$ & 0.026 & -- \\
  Ground matrix  & $\phi^0_\text{g}$ & 0.274 & -- \\
\bottomrule
\end{tabular}
\end{table}

We quantify alterations at the tissue scale by tracking the evolution of the constituent partial densities $\hat{\rho}_i$ and the elastic stretches $\overline{\lambda}_i$, reported as the median across the LV domain.
We assess organ-level structural adaptations by evaluating the end-diastolic cavity volume, reported via the relative change from the initial no-growth state, $\Delta V_{\text{cav}}(\%)$, and the incremental change per G\&R step, $\Delta V_{\text{inc}}(\%)$.
To characterize the VO eccentric G\&R phenotype, we evaluate the three-dimensional sphericity index $\text{SI}$ \citep{Zeng2017}, which quantifies dilation toward a spherical geometry \citep{Opie2006} and is defined as:
\begin{equation}
\text{SI}  = \frac{\text{EDV}}{\tfrac{4}{3}\pi r^{3}} \, ,
\end{equation}
where $r$ is the LV long-axis half length.

Lastly, to quantify G\&R-induced changes in myocardial tissue stiffness, we evaluate the stress response of a virtual myocardial specimen under equibiaxial tension. 
The specimen is defined in the material plane spanned by the fiber direction $\vec{f}$ and the sheet-normal direction $\vec{n}$, with equal principal stretches prescribed along these two orthogonal directions. 
For simplicity, we assume a fully incompressible response. 
At each G\&R time point $\tau$, the specimen is assigned the constituent densities predicted by the ventricular simulation, taken as median values across the LV domain. 
This virtual test therefore isolates the effect of evolving tissue composition on the mixture-level constitutive response, while excluding changes associated with ventricular geometry or the in vivo loaded baseline state. 
Further details on the analytical derivation of the biaxial tensile test are provided in \hyperref[sec:appendix_c]{Appendix C}.

\section{Numerical results} 
\label{sec:numresults}
We assess the individual and combined contributions of collagen remodeling and myocyte hypertrophy on local tissue compliance and global LV adaptation by considering three scenarios: 
\textbf{(C) collagen-only}, where we prescribe a collagen mass loss at fixed myocyte mass; 
\textbf{(M) myocyte-only}, where we prescribe myocyte growth at constant collagen mass, and
\textbf{(C+M) combined} integrating both effects. 
The first two scenarios assess the sensitivity of the multi-constituent framework to the minimum collagen mass ratio $\hat{m}_\text{c,min}$ and the myocyte growth rate constant $\kappa_\text{m}$, while the combined case captures their joint influence on the LV structure and function.

\subsection{Collagen-driven adaptation}
\label{Ccase}
This section presents the \textbf{(C) collagen-only} scenario, where collagen mass gradually decreases, while myocyte mass remains constant.

\begin{figure*}[h!]
\centering
\includegraphics[width=\textwidth]{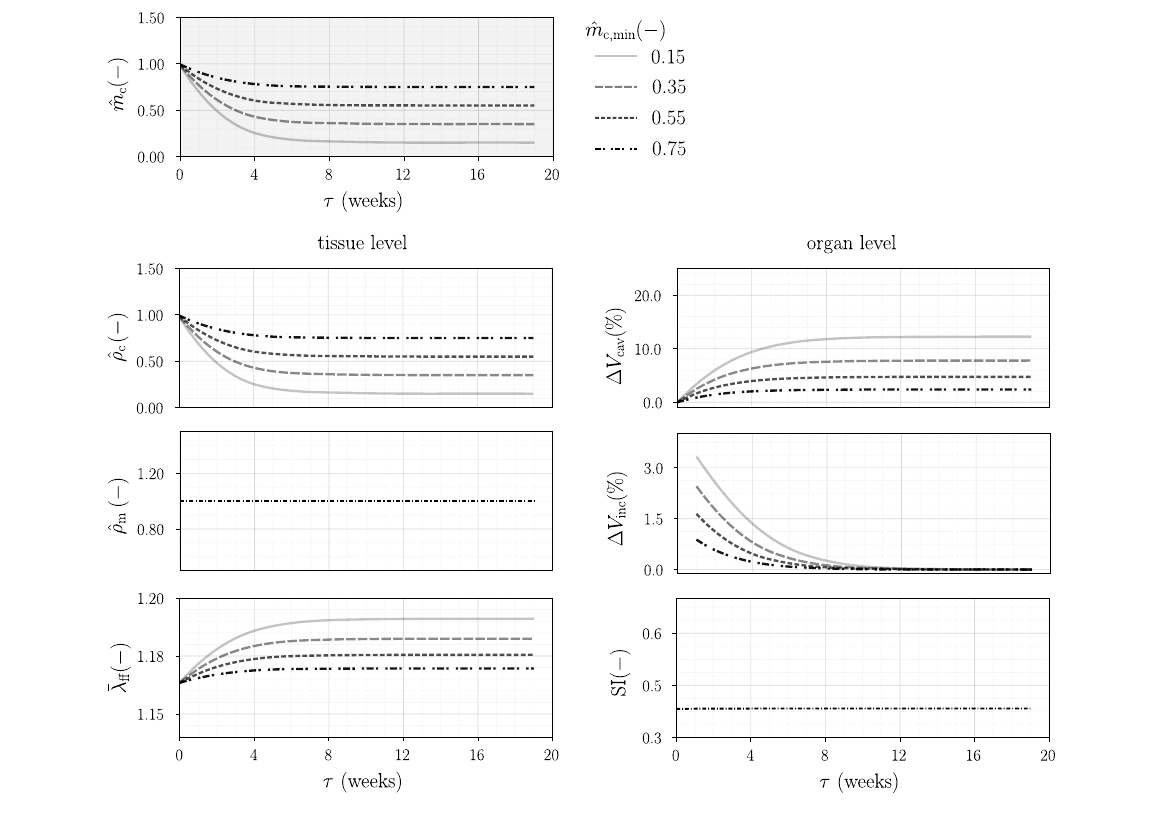}
    \caption{\textbf{(C) collagen-only -- Local and global changes.} In response to six weeks of collagen mass decay with four different minimal mass ratio $\hat{m}_{\text{c,min}} = [0.15, 0.35, 0.55, 0.75]$, G\&R effects are quantified over 19 weeks. Week 0 is the baseline and corresponds to the reference no-growth configuration. On the left, tissue-level changes are represented through the evolution of collagen partial density $\hat{\rho}_\text{c}$, myocyte partial density $\hat{\rho}_\text{m}$, and the local elastic stretch along the myofibers $\overline{\lambda}_\text{ff}$. On the right, global changes reflect alterations at the organ level, quantified through the relative $\Delta V_\text{cav}$ and the incremental $\Delta V_\text{inc}$ change in LV cavity volume, and the sphericity index $\text{SI}$}.
\label{fig:collagen-driven}
\end{figure*}

In subjects undergoing persistent VO, collagen degradation typically stabilizes after four to six weeks, with no further changes thereafter \citep{Corporan2018}.
We represent collagen degradation by prescribing a decay in its mass, see Eq.\eqref{eq:coll_degrad}, with $\tau_\text{end} = 6$ weeks and vary the minimum mass ratio $\hat{m}_{\text{c,min}} = [0.15, 0.35, 0.55, 0.75]$ to assess the sensitivity to low and high mass loss. 
Figure \ref{fig:collagen-driven} summarizes collagen-driven LV adaptations in response to a prescribed collagen mass loss, as shown in the top gray graph.
Resulting modifications at the tissue scale are shown on the left of the figure: change in myocyte partial density $\hat{\rho}_\text{m}$ (first plot), in collagen partial density $\hat{\rho}_\text{c}$ (second plot), and in elastic stretch $\bar{\lambda}_\text{ff}$ along the fiber direction (third plot).
Changes at the organ level are reported on the right of the figure: evolution of the relative $\Delta V_\text{cav}$ (first plot) and the incremental $\Delta V_\text{inc}$ (second plot) change in LV cavity volume, and of the sphericity index $SI$ (third plot).
Note that myocyte and collagen fibers aligned with their principal direction experience an elastic stretch $\bar{\lambda}_\text{m}$ = $\bar{\lambda}_\text{cf}$ = $\bar{\lambda}_\text{ff}$. 

\textbf{Local changes at the tissue scale.}
In a pure DA scenario, there is no volumetric growth, i.e., $\ten{F}^\text{g} = \ten{I}$ and $\det\ten{F}^\text{g} = \hat{v} = 1$. 
Consequently, the change in collagen partial density matches the change in collagen mass.
In contrast, myocyte partial density remains unchanged. 
Locally, the elastic fiber stretch initially rises during the first six weeks, and then plateaus once collagen degradation reaches its final value. 

\begin{figure*}[h]
\centering
\includegraphics[width=\textwidth]{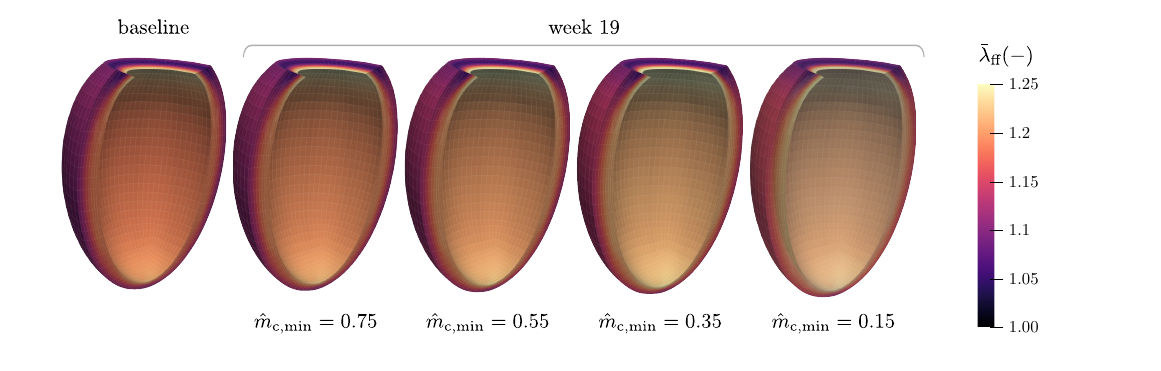}
\caption{\textbf{(C) collagen-only -- Elastic stretch.} The spatial distribution of the elastic stretch along the mean fiber direction is shown at Week 19, at the end of the simulated growth. Collagen mass loss yields higher elastic stretches, which increase from the endocardium to the epicardium. The increase in EDV volume does not produce a significant shift from an elliptic to a spherical phenotype.}
\label{fig:collagen-driven-stretch}
\end{figure*}

\textbf{Global changes at the organ scale.}
Changes in the tissue microstructure can affect the loaded ventricular geometry at the organ level. 
Globally, an increase in the end-diastolic stretch translates to a higher EDV. 
In response to VO, the LV cavity volume increases from 2.4 to 12.3\%, corresponding to 25 and 85\% of collagen degradation, respectively, during the first 19 weeks of compensated VO-induced G\&R. 
The sphericity index remains unaltered, with $\text{SI} \sim 0.4$. The results suggest that homogeneous collagen degradation across all fiber orientations does not drive any appreciable change in LV phenotype towards an eccentric pattern.
Following the course of collagen degradation, most significant changes occur within the first six weeks, with the incremental change in cavity volume reaching a maximum value of 3.3\% in the first week for $\hat{m}_\text{c,min}=0.15$.  
The resulting LV geometry at the end of G\&R in Week 19 is provided in Fig. \ref{fig:collagen-driven-stretch}. 
Elastic stretches vary transmurally, showing higher values at the endocardial surface. Their peak value increases with larger collagen mass loss.

\begin{figure*}[h]
\centering
\includegraphics[width=\textwidth]{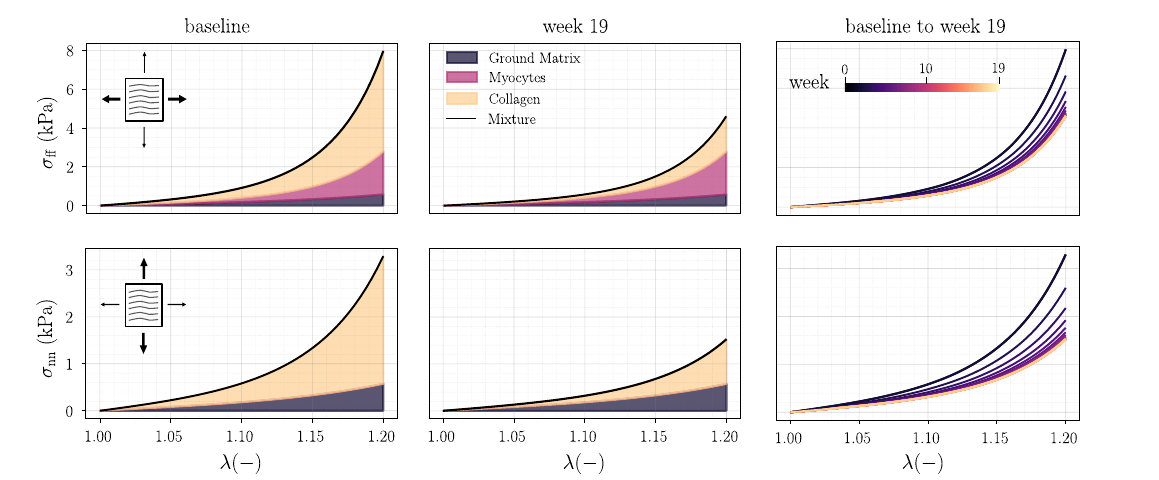}
\caption{\textbf{(C) collagen-only -- Evolving tissue compliance.} The stress-stretch behavior along the fiber and sheet-normal direction is plotted at baseline (left) and at the end of Week 19 (middle) for $\hat{m}_{\text{c,min}} = 0.35$. On the right, the total biaxial stress is computed at each G\&R time step to capture the evolution of the material behavior. Each curve refers to one G\&R week, spanning from the baseline no-growth time to the end of the simulated growth in Week 19.}
\label{fig:collagen-driven-biaxial}
\end{figure*}
\textbf{Evolving tissue compliance.}
To assess the impact of pure collagen-DA myocardial G\&R on tissue compliance, Figure \ref{fig:collagen-driven-biaxial} (middle) shows the biaxial stress-stretch response along the myofiber and the cross-myofiber direction, at the end of Week 19 for $\hat{m}_{\text{c,min}} = 0.35$.
At baseline, collagen fibers contribute significantly to the stiffness in the fiber direction and dominate the mechanical behavior in the sheet-normal direction.
Compared to the baseline, the myocardial stiffness decreases in both directions, as reflected by the downward shift of the stress-stretch curve.
Figure \ref{fig:collagen-driven-biaxial} (right) illustrates how collagen degradation alters the load-bearing behavior throughout the simulated G\&R period. 
The first weeks exhibit the largest shift, directly following the rapid degradation in collagen mass over the first six weeks.                   
In summary, the \textbf{(C) collagen-only} model indicates that early collagen mass degradation, occurring mainly within the first six weeks following VO induction, substantially increases tissue compliance. As a consequence, such a reduction in load-bearing capacity of the tissue promotes LV dilatation even in the absence of volumetric growth.

\subsection{Myocyte-driven adaptation}
\label{Mcase}
\begin{figure*}[h]
\centering
    \includegraphics[width=\textwidth]{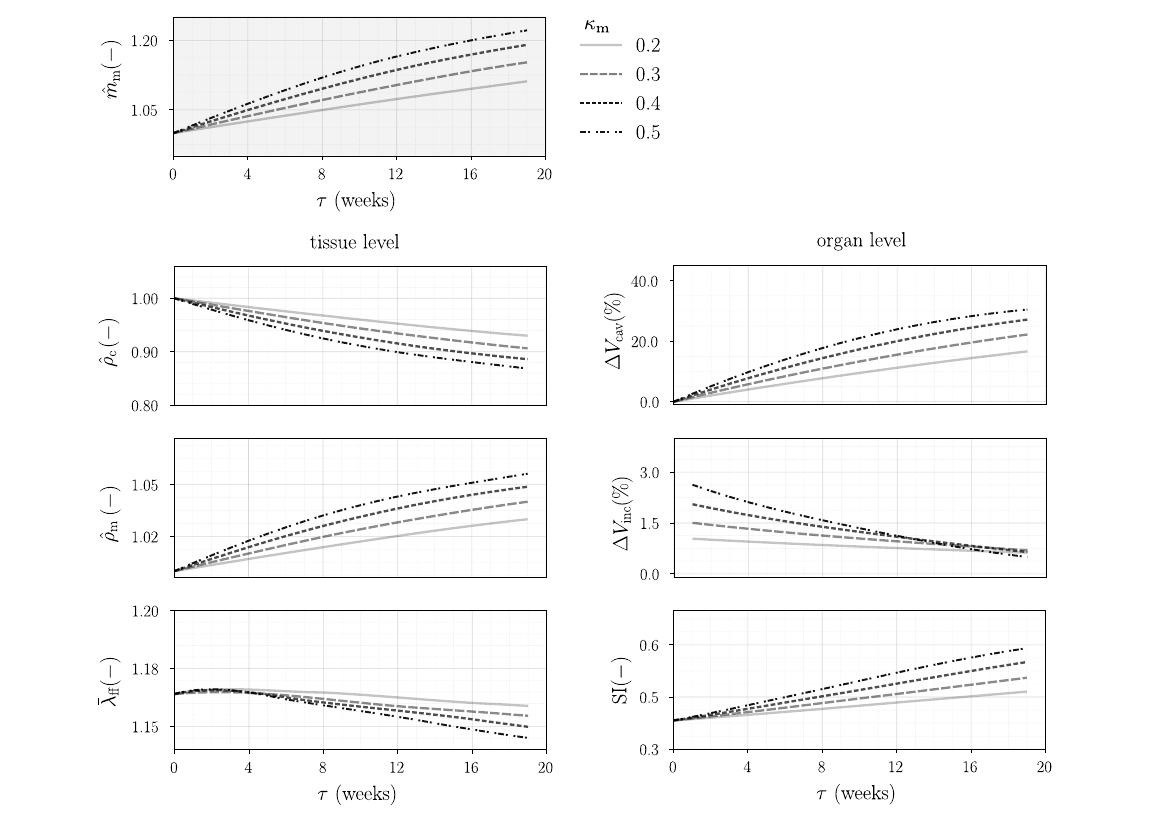}
        \caption{\textbf{(M) myocyte-only -- Local and global changes.} G\&R changes are quantified over 19 weeks for four myocyte growth rates $\kappa_m = [0.2, 0.3, 0.4, 0.5]$. On the left, tissue-level changes are represented through the evolution of the collagen partial density $\hat{\rho}_\text{c}$, the myocyte partial density $\hat{\rho}_\text{m}$, and the local elastic stretch along the myocyte direction $\overline{\lambda}_\text{ff}$. Global changes reflect alterations in the left ventricular cavity volume, quantified through the relative $\Delta V_\text{cav}$ and the incremental $\Delta V_\text{inc}$ change in LV cavity volume, and the sphericity index $\text{SI}$}.  
    \label{fig:myocyte-driven}
\end{figure*}

This section focuses on the \textbf{(M) myocyte-only} case for which myocyte growth in response to VO is stretch-driven. Thus, myocyte mass evolves as long as the elastic fiber stretch surpasses the homeostatic state. 
Following a pure VA scenario, collagen mass is kept unchanged.
Myocyte-driven adaptation is modeled through the rate-based law in Eq.\eqref{eq: stimulus myocytes}, where we vary the growth rate parameter $\kappa_\text{m} = [0.2,\, 0.3,\, 0.4,\, 0.5]$ weeks$^{-1}$.  
Similarly to Fig. \ref{fig:collagen-driven}, Figure \ref{fig:myocyte-driven} summarizes the evolution of local and global quantities in response to VO.  

\textbf{Local changes at the tissue scale}. 
Following the VA assumption, local changes in myocyte mass happen through volumetric growth, i.e., $\ten{F}^\text{g} \neq \ten{I}$ and $\det \ten{F}^\text{g} = \hat{v} \neq 1$. 
In our model, we specify an anisotropic growth deformation gradient via Eq.\eqref{eq: transv_growth} and the additional volume is accommodated along the principal myofiber direction. 
In response to VO, myocytes increase their mass during the 19-week G\&R period, reaching $1.11$ and $1.22$ times their initial mass for the extreme cases with $\kappa_\text{m} = 0.2$ and $0.5$, respectively.
Myocyte elongation of 9\% and 18\% was measured experimentally at Week 9 and 16 \cite{Ryan2007}, which is in close agreement with our result for $\kappa_\text{m} = 0.4$. 
The volumetric change $\hat{v}$ leads to updated partial densities for both myocyte and collagen, and the collagen partial density decreases as a consequence of myocyte growth in volume. 
A key distinction with the \textbf{(C) collagen-only} case is that the elastic fiber stretch gradually decreases during G\&R instead of reaching a plateau. 
According to the growth kinematics, myocytes elongate so as to restore the homeostatic state and meet the elastic homeostatic stretch. 
Without any change in tissue properties, this inelastic growth progresses at a pace fixed by the growth rate parameter until the target homeostatic stretch is reached at each material point.
The higher the growth rate parameter, the faster the mechanical stimulus evolves towards restoring its homeostatic value.

\begin{figure*}[h]
\centering
\includegraphics[width=\textwidth]{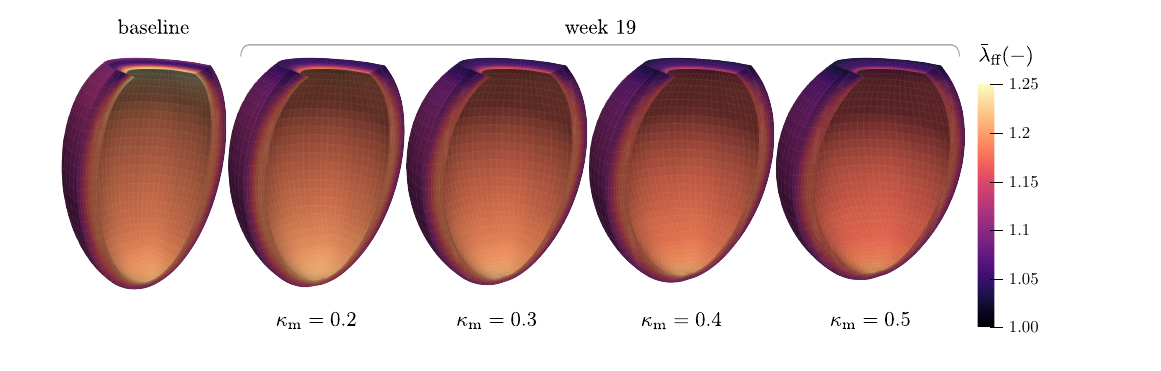}
\caption{\textbf{(M) myocyte-only -- Elastic stretches.} The spatial distribution of the elastic stretch along the fiber direction is shown at Week 19. Myocyte mass growth lowers elastic stretches in the LV. At the organ scale, it promotes eccentric growth and a more spherical LV shape.} 
\label{fig:myocyte-driven-stretch}
\end{figure*}

\textbf{Global changes at the organ scale.}
At the end of Week 19, myocyte hypertrophy induces an increase in EDV cavity of 17\% and 27\% for the extreme cases with $\kappa_\text{m} = 0.2$ and $0.5$, respectively. 
Similar to the \textbf{(C) collagen-only} scenario, most changes occur before the first ten weeks of G\&R. 
In fact, the mechanical stimulus felt by myocytes is progressively reduced as the deviation from the homeostatic state decreases. 
Accordingly, the incremental change in cavity volume decreases from 2.4\% to 0.5\% from Week 3 to 19 for $\kappa_\text{m} = 0.5$. 
The geometric modification induced by myocyte G\&R results in the eccentric phenotype, where the LV assumes a more spherical shape \citep{Zeng2017}. 
This is reflected by the resulting sphericity index, which increases from $\text{SI} = 0.38$ to a maximum of $0.6$ at Week 19. 
Fig. \ref{fig:myocyte-driven-stretch} visualizes the distribution of the elastic fiber stretch acting as a mechanical stimulus for growth at Week 19. 
The LV endocardium exhibits the largest stretches due to intraventricular pressure loading, with maximum values occurring in proximity to the LV apex.

\begin{figure*}[h]
\centering
\includegraphics[width=\textwidth]{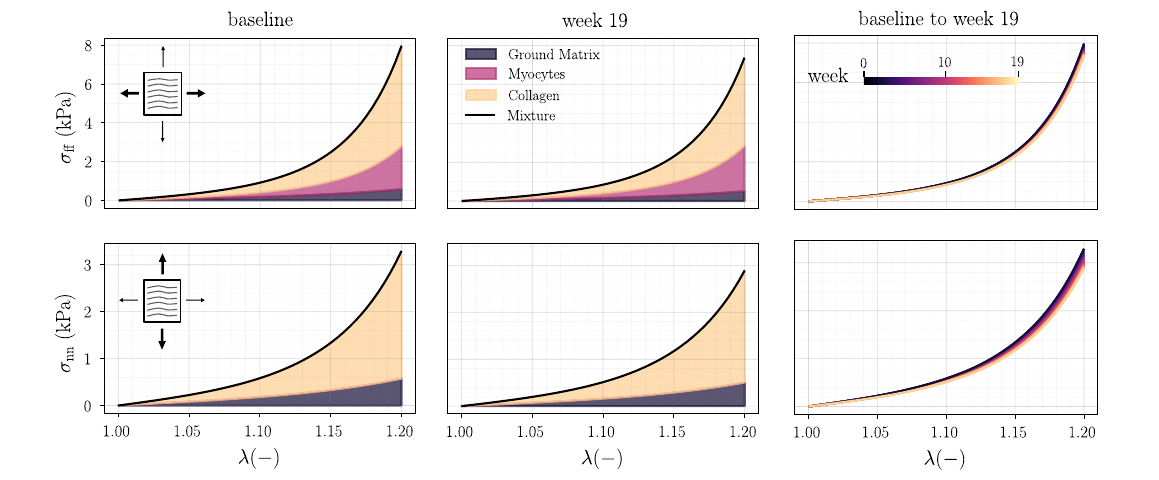}
\caption{\textbf{(M) myocyte-only -- Evolving tissue compliance.} Stress–stretch responses along the fiber and sheet-normal directions are plotted at baseline (left) and Week 19 (middle) for $\kappa_\text{m} = 0.4$. The right graph shows the total biaxial stress computed at each G\&R time step, where each curve refers to one G\&R week.} 
\label{fig:myocytes-driven-biaxial}
\end{figure*}

\textbf{Evolving tissue compliance.}
The biaxial tensile test, performed at the end of each G\&R week, shows almost no change in the stress-stretch behavior of the myocardium, as illustrated in Fig. \ref{fig:myocytes-driven-biaxial} for $\kappa_\text{m} = 0.4$. 
VA of myocytes does not directly affect the overall tissue constitutive response. 
However, the increase in volume by myocyte hypertrophy lowers the partial density of collagen. 
Consequently, snapshots of the constituent contributions to the tissue stiffness at Week 0 and 19, see Fig. \ref{fig:myocytes-driven-biaxial} (left and middle), show that the biaxial mechanical response exhibits a slight softening in the sheet-normal direction, where collagen is the primary load-bearer. 

In summary, our \textbf{(M) myocyte-only} model shows that myocytes contribute to LV G\&R by increasing their mass in an attempt to bring the elevated stretch induced by VO back to their homeostatic value. 
At the organ scale, myocyte hypertrophy drives the enlargement of the LV, producing the eccentric phenotype, with only marginal changes to the tissue mechanical properties.

\subsection{Combined collagen and myocyte adaptation}
\label{CMcase}
\begin{figure*}[h!]
    \includegraphics[width=\textwidth]{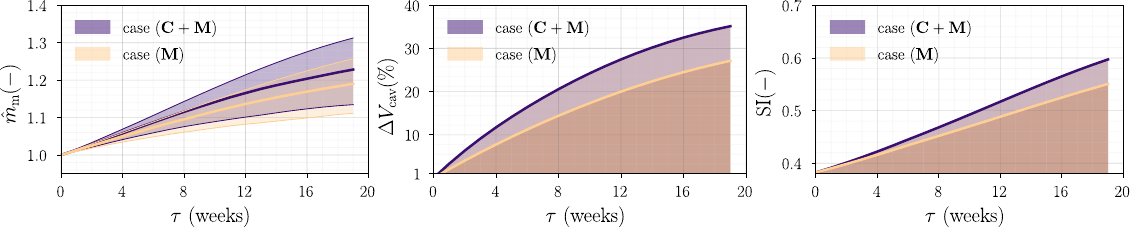}
        \caption{\textbf{(C+M) combined collagen and myocyte -- global and local changes.} G\&R changes are quantified over 19 weeks for the \textbf{(C+M) combined}, and compared to the \textbf{(M) myocyte-only} case. On the left, we show the evolution of myocyte partial mass $\hat{m}_\text{m}$. Changes in cavity volume $\Delta V_\text{cav}$ and sphericity $\text{SI}$ are shown in the middle and on the right, respectively, quantifying organ-level alterations.}
    \label{fig:collagen and myocyte-driven}
\end{figure*}

In previous subsections, we isolated the contributions of myocyte and collagen, and carried out a sensitivity analysis of the G\&R parameters. 
We now reproduce a physiologically relevant case, referred to as the \textbf{(C+M) combined} case, to study the reciprocal roles of collagen and myocyte adaptation during compensatory stages post-VO. 
We set $\kappa_\text{m}=0.4$, which in the \textbf{(M) myocyte-only} scenario yielded the closest match to the hypertrophy measured experimentally \citep{Ryan2007}. 
Simultaneously, we prescribe $\hat{m}_\text{c,min} = 0.7$ and $\tau_\text{end} = 5$ week, leading to an experimentally observed collagen mass degradation of 30\% at Week 5 \cite{Ryan2007}, which is maintained unaltered for the remaining G\&R time. 
In Fig. \ref{fig:collagen and myocyte-driven}, local and global changes associated with this \textbf{(C+M) combined} scenario are presented along a pure \textbf{(M) myocyte-only} scenario based on the same growth parameters. 
In the following, changes at the tissue are quantified through the evolution of myocyte partial mass, and at the organ scales by changes in LV volume and sphericity.
    
\textbf{Local changes at the tissue scale}. %
The presence of 30\% collagen degradation, albeit moderate, further enhances myocyte hypertrophy, with myocyte mass increasing from 6.2\% and 16\% in the \textbf{(M) myocyte-only} scenario and from 7.3\% and 19\% in the \textbf{(C+M) combined} scenario at Weeks 8 and 15 post-VO, respectively. 
In fact, the decrease in collagen mass density during the first six weeks alters the overall mechanical behavior of the tissue by increasing its compliance and, in turn, the end-diastolic elastic stretches, as already shown in Subsec. \ref{Ccase} .
Since myocyte G\&R is mechanically driven, higher stretches lead to greater increase in myocyte mass, see Fig. \ref{fig:collagen and myocyte-driven} (left), in the \textbf{(C+M) combined} scenario compared to the \textbf{(M) myocyte-only} scenario. 

\begin{figure*}[h!] 
\centering
\includegraphics[width=\textwidth]{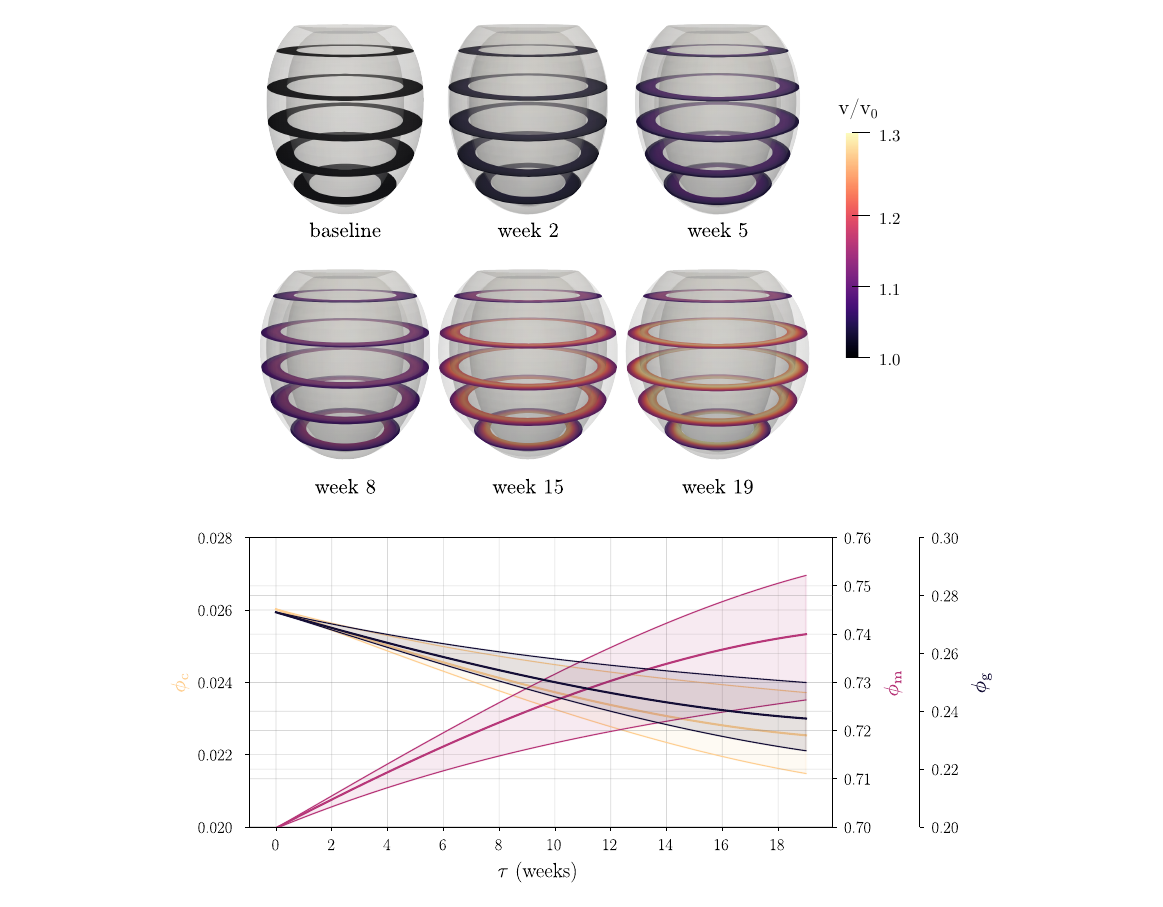}
\caption{\textbf{(C+M) combined collagen and myocyte -- Volume change and volume fractions.} The total tissue volume change is shown at Week 0 (baseline), 5, 8, 15, and 19 (top). The local evolution of volume fractions is shown for myocardial constituents (myocytes, collagen, ground matrix). Thick solid lines represent median values, while shaded areas indicate the interquartile range (25th–75th percentiles) across the whole LV domain (bottom). } 
\label{fig:total volume and vol fractions}
\end{figure*}

Thus, early collagen degradation not only promotes ventricular dilatation by reducing myocardial stiffness but also enhances myocyte hypertrophy.
Figure \ref{fig:total volume and vol fractions} (first plot) illustrates the local volume change driven by myocyte and collagen mass changes at Weeks 0, 5, 8, 15, and 19 post-VO onset. Myocyte hypertrophy induces heterogeneous volumetric growth across the ventricular wall, characterized by elevated growth at the endocardium, mirroring the transmural distribution of myocardial strain. By the end of Week 19, the tissue volume increases by approximately 30\% relative to its initial value.
After each G\&R cycle, individual volume fractions are updated, reflecting changes in the volume occupied by each constituent with respect to the total tissue volume, see Fig. \ref{fig:total volume and vol fractions} (second plot). 
Consistent with mass changes in both collagen and myocytes, myocyte volume fraction increases, whereas both collagen and ground matrix fractions decrease throughout G\&R.

\textbf{Global changes at the tissue scale}. 
Myocyte growth alone in the \textbf{(M) myocyte-only} scenario results in a 27\% increase in LV volume by Week 19.
When myocyte growth is accompanied by collagen degradation in the \textbf{(C+M) combined} scenario, the cavity volume increases by 35\%, corresponding to an LV volume of 194\,ml at Week 19, see \ref{fig:collagen and myocyte-driven} (middle). 
Collagen loss, by promoting the hypertrophic response of myocytes, favors geometrical and structural changes in the LV, which ultimately leads to higher ventricular enlargement. 
This is also reflected by the G\&R phenotype of the LV, as the transition toward a more spherical chamber geometry is further accentuated, with the sphericity index reaching $\text{SI}=0.55$ in the \textbf{(M) myocyte-only } scenario compared to $0.6$ in the \textbf{(C+M) combined} scenario.

In summary, the reciprocal role of collagen degradation and myocyte growth can be explained mechanistically with the \textbf{(C+M) combined} scenario, such that collagen loss causes an increase in myocardial compliance, resulting in increased end-diastolic stretches, i.e., a higher mechanical stimulus for myocyte hypertrophy. Globally, this combined G\&R results in a more dilated and spherical LV. 

\section{Discussion}
\label{sec:discussion}
Hearts subjected to VO experience a series of distinct anatomical and functional modifications, which include LV dilatation and increased sphericity on the one hand, and alterations in tissue compliance on the other hand \citep{Bonow1992, Melenovsky2013, Ahmed2022InterstitialRegurgitation}. 
Such macroscopic changes are driven by underlying microstructural adaptations of the tissue constituents, predominantly involving myocyte growth and collagen remodeling.
While myocyte growth has been largely studied both numerically and experimentally and is recognized as the dominant cellular event contributing to LV eccentric hypertrophy \citep{Emery1997, Wisdom2015, Sharifi2021, Peirlinck2019}, the intrinsic role of collagen remodeling in this evolving process is often overlooked. 
Although experimental studies suggest that both collagen-driven alterations in material properties and myocyte–driven hypertrophy are central factors in the progression of VO \citep{Spinale2000, Spinale2007MyocardialFunction, Ryan2007, Corporan2018}, the cause-and-effect relationships governing the interplay between these processes remain vastly underexplored.
In this contribution, we integrated a mixture-based material description within the kinematic growth framework, in order to couple mechanically stimulated mass changes in myocytes, collagen, and ground matrix to alterations in material behavior and in volumetric growth.
Our formulation allows myocytes to vary their mass through VA, while collagen is assumed to follow DA. 
Within an idealized human LV model, we first isolated the 
contributions of myocyte growth and collagen remodeling, and assessed their dominant effects on the organ-level phenotype and the material behavior. Secondly, we examined their combined influence, demonstrating the synergy between myocyte growth and collagen remodeling, which together drive both constitutive and volumetric changes. 

\subsection{Collagen degradation and myocardial softening}
The developed G\&R framework enables relating mass changes in individual tissue constituents to both organ-scale geometrical changes, such as alterations in LV dimensions and phenotype, and tissue-scale constitutive changes, reflected by the shift in the tissue stress-strain curve.
The first objective is to discern which of these alterations are predominantly driven by myocyte or collagen G\&R (and to what extent), thereby isolating the respective roles of these processes in governing LV adaptation in response to VO. 
Our numerical results indicate that collagen degradation alone alters myocardial tissue properties by inducing a softening in the stress-stretch response during the early (15-20 weeks) compensatory progression of VO-induced G\&R. 
As changes in collagen mass through DA do not yield volumetric growth, elastic stretches accumulate in the tissue without any compensatory inelastic deformation.
At the organ level, reduced load-bearing capacity leads to ventricular dilatation during G\&R \citep{Corporan2018}.
Similar results \citep{Wang2016Image-drivenFibrosis} were obtained for myocardial infarction and linked constitutive changes to collagen mass alteration. 
Interestingly, their findings suggested that collagen fibrosis, i.e., an increase in collagen mass, reduces myocardial compliance. The increase in stiffness prevents the LV from properly relaxing and filling, thereby promoting diastolic dysfunction \citep{Conrad1995, Tanaka1986QuantitativeCardiomyopathy}. 
Notably, their image-based framework models fibrosis by modifying the strain energy function through weighting factors that directly represent variations in collagen fiber mass and thickness. These weights, once identified, are kept fixed to reproduce single-time-point ex vivo mechanical tests of infarcted tissue. Fibrosis is thus modeled at a single time point (end-stage), rather than by tracking its evolution over time.
Our G\&R framework represents myocardial adaptation by allowing the strain energy function to evolve in time through varying constituent partial densities and naturally captures the dynamics of G\&R processes.
In the context of cardiovascular tissue, a recent work by \citep{Sesa2025} developed an energy-based model for the evolution of collagen density and reorientation 
during maturation, where strain energy functions depend explicitly on collagen density.
Similarly to our formulation, they coupled the constitutive description with volumetric growth. Instead of prescribing the growth deformation gradient tensor, a stress-driven homeostatic surface approach was adopted. In their model, all constituent densities remain constant, while only collagen density is set to change, without capturing the mutual influence among constituents. In our study, individual constituent changes in mass and volume mutually affect their relative amounts in the tissue through partial mass densities, which results in the redistribution of load among structural constituents.

\subsection{Myocyte hypertrophy and eccentric LV phenotype}
When simulating purely myocyte-driven G\&R, myocyte hypertrophy produces a marked increase in the LV cavity volume, reaching 27\% above its baseline dimensions in response to a 22\% increase in myocyte mass. 
The inelastic growth induced by myocytes acts to restore the elastic stretch toward its homeostatic value. 
Our G\&R model captures characteristic features of myocyte-driven eccentric growth in the LV: myocyte sarcomerogenesis and elongation promote chamber dilatation with added volume depositing along the fiber direction \citep{Gktepe2010, Genet2016, Peirlinck2019}. 
Consistently with previous findings \citep{Kroon2009, Peirlinck2019}, our models shows that hypertrophy initiates in the endocardium, where on average higher elastic stretches are found, and progresses towards the epicardium. 
Our numerical results show a progressive transition of the LV from an elliptical to a more spherical shape, with larger values of the sphericity index associated with increased myocyte growth. 
A similar trend was reported by \citep{Genet2016}, who qualitatively observed increased sphericity in their eccentric G\&R simulations. 
Under the VA assumption, myocyte hypertrophy increases tissue volume, which indirectly marginally alters the material properties, i.e., a slight tissue softening due to reduced collagen partial densities. 

\subsection{Interplay between collagen and myocytes: synergy and compensation}
The relevance of the proposed framework becomes evident when tackling a physiologically realistic scenario, in which organ-level adaptation does not arise from a single process, such as myocyte hypertrophy or collagen degradation, but rather from the interplay between the G\&R pathways of individual constituents. 
Our model captures both structural and functional ventricular changes resulting from the alterations in collagen density and myocyte volume under VO. 
In the presence of both myocyte hypertrophy and collagen degradation, the reduction in collagen mass density increases tissue compliance and end-diastolic stretch, thereby amplifying the mechano-driven myocyte growth response, leading to a larger increase in cavity volume and an accentuated eccentric growth. 
Besides the present work, the relation between microstructure and function in hearts undergoing VO 
has been explored computationally only by \citep{Guan2023}. In this study, the authors coupled a constitutive mixture model to kinematic growth within an evolving reference configuration approach. Similarly to our approach, the constitutive model explicitly accounts for myocardial tissue constituents in the strain energy function, weighted by relative volume fractions. In modeling G\&R, the evolution laws rely on a single growth multiplier, which is either applied (i) to both collagen and myocytes uniformly, or (ii) to myocytes only. As a result, their model suggests that collagen either increases in volume fraction, leading to a progressively stiffer stress–strain response, or exhibits only minimal decreases in volume fraction as a secondary effect of myocyte growth. 
Conversely, our framework assigns distinct G\&R process dynamics and prescribes a specific growth type  
for each constituent, based on experimental observations. This distinction allows for differentiating between structural changes driven by myocyte elongation and those resulting from collagen degradation. Beyond VO studies, cardiac G\&R has recently been investigated using homogenized CM theory in the context of pressure overload 
induced by hypertension \citep{Gebauer2023}. Here, the authors model G\&R through a continuous turnover of collagen and myocytes, which causes an inelastic remodeling. Our formulation differs in the following aspects. 
First, they model mass growth as an elastic swelling, in which added mass deposits in the direction that ensures constant spatial density. We follow, instead, the eccentric growth hypothesis, and prescribe the inelastic growth deformation gradient along the longitudinal axis of the myocyte \citep{Goriely2017, SahliCostabal2019MultiscaleFailure}. 
Second, rather than accounting for mass growth collectively for all constituents, we model volume and density growth separately for each constituent \citep{Schmid2012}. 
As such, the total tissue volumetric change is computed as a weighted sum of the constituent contributions, using their initial volume fractions. Consequently, myocytes, which constitute approximately 70\% of the initial tissue volume, act as the main driver for volumetric growth.

\subsection{Comparison to experimental studies}
Altogether, our results demonstrate that although myocyte mass growth is the predominant driver for ventricular dilatation at the organ level, collagen degradation not only accommodates dilatation but also promotes further myocyte hypertrophy. 
Experimentally, rodent and large animal models suggest that interstitial collagen loss exacerbates myocyte elongation, irrespective of the mode of induction, e.g., aortocaval fistula and mitral regurgitation \citep{Perry2002AngiotensinDog, Spinale2007MyocardialFunction, Ryan2007, Corporan2018}. 
In their work, \citep{Ryan2007} combined LV chamber, cardiomyocyte, and interstitial collagen analyses to assess myocardial changes in aortocaval fistula-induced VO. 
They quantified chronic collagen degradation occurring in the first five weeks of 30\% loss and measured that
a myocyte elongation of 9\% and 18\% at the eight and 15 weeks after VO induction.  
Furthermore, they demonstrated that early collagen degradation, within the first weeks post-VO, is the primary G\&R mechanism for ventricular dilatation at an early stage, while myocyte adaptation is secondary. 
Our results are in good agreement with the chronic myocyte and collagen mass changes reported in the aforementioned study.
Notably, \cite{Ryan2007} reported an acute significant dilatation within the first week, $20\%$ change in end-systolic dimensions, attributed solely to collagen degradation. 
Contrarily, our model predicts that chronic collagen degradation contributes little to ventricular dilatation. 
However, our simulations neglect the acute collagen decrease of 75\% that occurs immediately after the onset of VO, within the first days, and instead focus on chronic adaptation over weeks. 
This explains that, compared to their findings, we observe reduced end-diastolic cavity volumes during the early weeks.
To date, no longitudinal studies have examined the relationship between myocyte and collagen sub-processes induced by VO in humans. 
Recently, \citep{Ahmed2022InterstitialRegurgitation} performed cardiac magnetic resonance in a cohort of 55 patients, complemented by biopsies obtained during mitral valve surgery. 
Their findings align with animal studies, confirming that loss of endomysial collagen is a root cause of increased compliance and ventricular sphericity. 
With these observations, we emphasize the need for longitudinal investigations that integrate histological and imaging techniques to quantify both collagen and myocyte mass changes and to assess their contributions to acute and chronic alterations in VO \cite{SahliCostabal2019MultiscaleFailure,Peirlinck2019}.

\subsection{Outlook}
Building on this work, we foresee relevant future extensions to improve the mechano-biological interpretability of our G\&R framework. 
First, we prescribe collagen net loss through a decay function for collagen mass density, which serves as a simplified phenomenological representation of the complex biochemical processes governing collagen turnover. In reality, these processes are mediated by dysregulation of matrix metalloproteinases (MMPs) and their inhibitors (TIMPs), favoring collagen degradation and resulting in diminished tissue stiffness \citep{Spinale2000}.
Recent numerical studies have incorporated key biochemical regulators of collagen remodeling, such as the fibrogenic factor TGF-$\beta$, which promotes collagen deposition, as well as MMPs and TIMPs, into micromechanical models of collagen adaptation \citep{Saez2013, Jin2011, Jia2019}. Our multi-constituent framework naturally lends itself to the integration of such mechano-chemical models of collagen G\&R, including the approach proposed by \citep{Saez2013}, which couples collagen mass density production and degradation to imbalances among ECM matrix proteins. Mass rate equations would herein be coupled to specific signaling pathways, e.g., fibrogenic proteins and MMPs–TIMPs cascades. Incorporating these chemical regulatory mechanisms would enable the investigation of collagen deposition and removal across the acute, chronic, and late stages of VO \citep{Corporan2018}.
Second, our current mixture material model is calibrated on healthy human myocardial test data \citep{Sommer2015}. 
As detailed ex vivo testing data for diseased tissue are lacking, there is a strong need for more in-depth explorations of the intrinsic microstructural organization and functional behavior of diseased myocardial tissue.
Such data could further inform an extended material model that incorporates richer microstructural features, including separate endomysial and perimysial collagen and changes in fiber thickness and orientation \citep{Nevo1989, Sacks2003, Wang2015, Cardona2025}. Furthermore, such model could feature micromechanical aspects, for example through a strain-energy function incorporating collagen fibers dispersion parameters, and will allow us to capture the effects of fiber and sheet dispersion on the mechanical behavior of the myocardium during G\&R \citep{Eriksson2013,Wang2015}.
Finally, while most VO G\&R theories focus on passive stimuli for growth evolution laws, it is likely that changes in the microstructure affects the tissue active response. 
Therefore, extending our framework to consider active stress laws \citep{Guccione1993, Witzenburg2017} would allow quantifying the impact that structural tissue changes have on the cardiac function throughout the full cardiac cycle.
\section{Conclusions}
\label{sec:conclusions}
This study proposes a mechanical G\&R framework that
integrates a mixture-based constitutive model of myocardial tissue with the
kinematic growth formulation. 
Unlike classic kinematic growth-based models, such a framework links changes in individual volume or density of constituents to both changes in mechanical properties and in organ geometry.
We use our framework to explore the impact of individual and combined adaptations of myocyte and collagen on tissue material properties changes and whole-organ geometrical alterations.
We quantify changes in tissue composition by tracking individual masses, densities, and volume fractions, and structural modifications by tracking LV end-diastolic volumes, their relative changes from the initial no-growth state, and sphericity.  
Interestingly, our framework shows that collagen degradation exacerbates myocyte hypertrophy by increasing end-diastolic stretches, confirming experimental observations on large animals and rodents.
This result highlights the synergistic myocyte and collagen interplay that accelerates the compensatory progression toward decompensation and eventually, systolic dysfunction. 
By integrating mechanistic descriptions of individual tissue constituents into the G\&R model, this work constitutes an important step toward a more comprehensive understanding of VO-induced cardiac G\&R. Incorporating micromechanical features and biochemical cues would further enable the study of coupled mechano–chemical–structural processes across scales, extending the analysis beyond isolated pathways to a level of complexity that remains to date inaccessible to experimental approaches. 

\section*{Acknowledgments}
M. Peirlinck acknowledges the support from the European Research Council within the European Union’s Horizon Europe research and innovation program (VITAL - Grant No. 101136728). We thank F. Gijsen for many valuable discussions, and A. Kewley for his insightful input and technical ingenuity.
%
\section{Ex vivo parameter optimization}
\label{sec:appendix_a}

Constitutive parameters are first calibrated on combined ex vivo biaxial and triaxial shear test experimental data \cite{Sommer2015} to achieve the best fit for the Holzapfel-Ogden microstructure-inspired material model \cite{Holzapfel2009}. 
The constitutive parameters are initialized from the calibrated myocardial response in \citep{Martonova2024}. Because the fiber-direction contribution represents the combined passive tensile response of myocytes and collagen aligned with the mean fiber direction, we partition this contribution between the myocyte term and the fiber-aligned collagen term. Specifically, we prescribe a fixed load-sharing factor $\eta$=0.30, such that 30\% of the passive fiber-direction stiffness is assigned to myocytes and 70\% to collagen. This is implemented by scaling the linear stiffness-like parameters of the two terms:
\begin{equation}
    a_m = \eta\, a_f,
    \qquad
    a_{cf} = (1-\eta)\, a_f.
\end{equation}

This partition is used as a constitutive modeling assumption to initialize constituent-specific load sharing. Future longitudinal datasets combining myocardial mechanics with histological measurements could be used to calibrate the load-bearing distribution among constituents \citep{Wang2015}.

\begin{figure*}
\centering
\includegraphics[width=\textwidth]{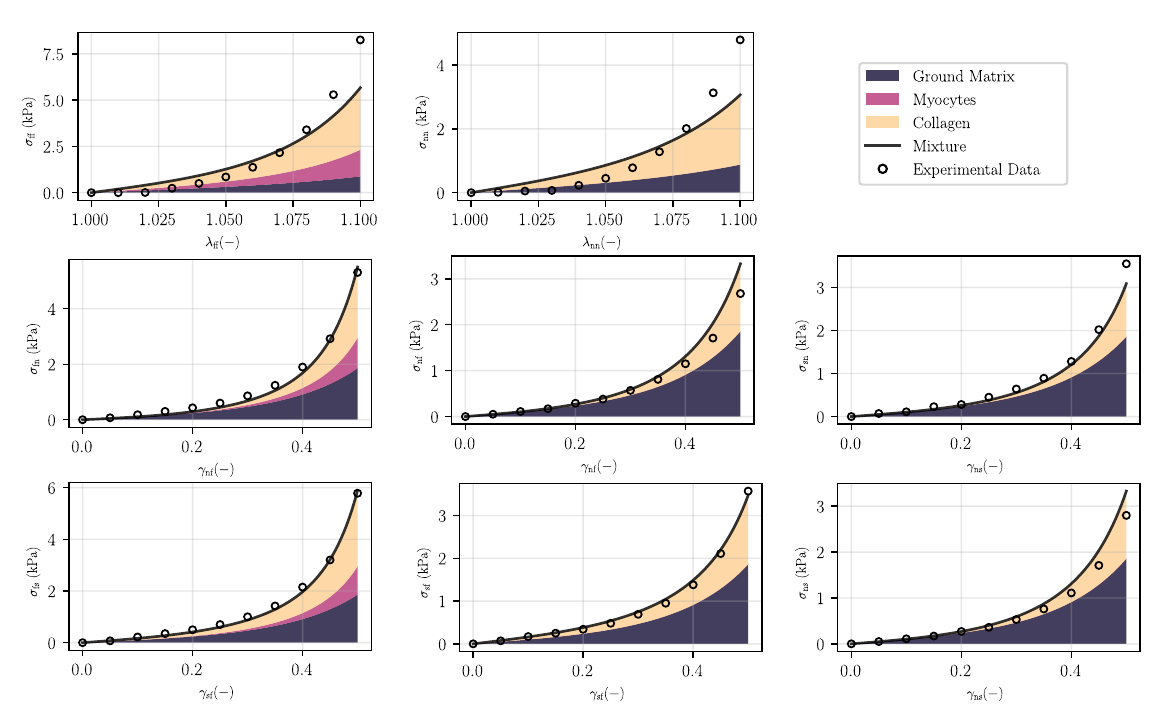}
\caption{\textbf{Ex vivo biaxial and shear behavior of the cardiac mixture}. Constitutive parameters of the cardiac mixture are adapted from \citep{Martonova2024} such that the best fit to ex vivo experimental data is achieved.
The contribution of each constituent in bearing the loading is displayed in terms of constituent-specific Cauchy stress (colored regions) as well as the total stress of the mixture (dark gray line). Experimental data are denoted with circles.} 
\label{fig:biaxial_ex_vivo}
\end{figure*}

Figure \ref{fig:biaxial_ex_vivo} shows the resulting stress responses for biaxial and triaxial shear tests along with the original experimental data.

\section*{Appendix B - In vivo parameter optimization}
\label{sec:appendix_b}
Constitutive parameters in Eq.~\eqref{eq: strain energy} are rescaled to reproduce a passive physiological response consistent with the in vivo end-diastolic pressure--volume relationship, following \citep{Peirlinck2018b}. This rescaling is motivated by the fact that constitutive parameters calibrated from ex vivo mechanical testing often overestimate myocardial stiffness under in vivo loading conditions, thereby requiring an additional calibration to ensure physiological behavior.
Linear coefficients \(a_\bullet\) and exponential coefficients \(b_\bullet\) are uniformly scaled by two scalar factors, \(A\) and \(B\), respectively.
The optimal values of \(A\) and \(B\) are determined by minimizing the least-squares (\(L^2\)-norm) error
between the simulated diastolic pressure--volume curve and the analytical Klotz curve \citep{Klotz2006DownloadedLib}. The objective function was evaluated over the volume range from the unloaded reference volume \(V_0\) to the end-diastolic volume at the prescribed end-diastolic pressure.  Minimal error ($e =0.23$) was achieved for $A =0.288$, $B = 0.393$.

\begin{figure}[h!]
\centering
\includegraphics[width=0.6\textwidth,clip=true,trim=4cm 0cm 4cm 1cm]{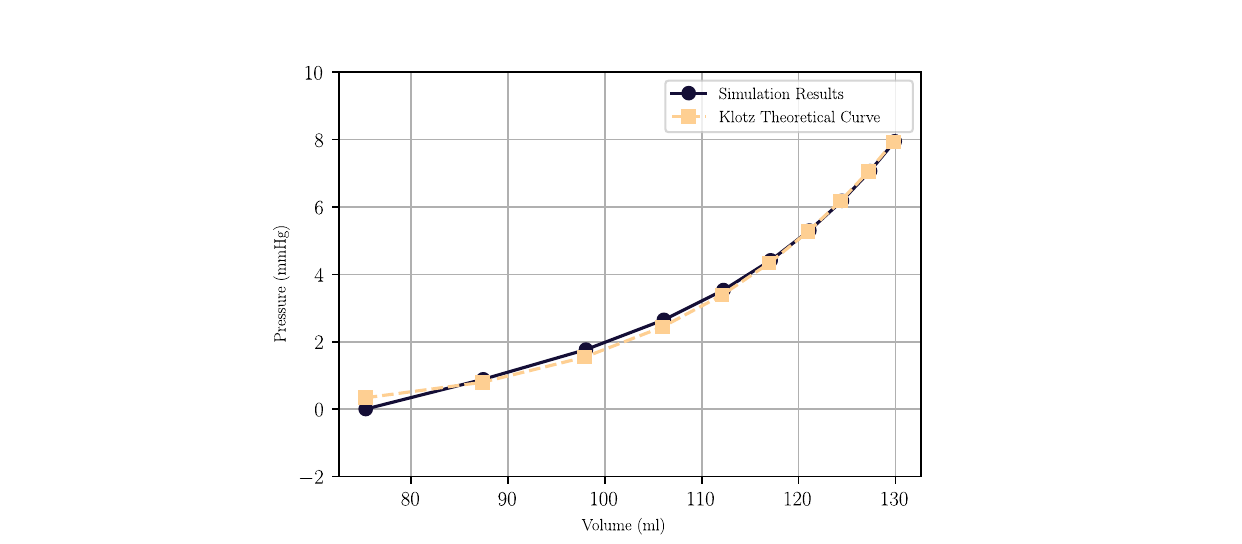}
\caption{\textbf{In vivo parameter optimization.}
Constitutive parameters are rescaled by minimizing the relative error between the simulated end diastolic pressure-volume curve (purple) and the theoretical Klotz curve (ochre) \citep{Klotz2006DownloadedLib}}.     
\label{fig:geometry_domain_boundaries_2}
\end{figure}

\section*{Appendix C - Evolving tissue compliance through a biaxial tension experiment}
\label{sec:appendix_c}
\setcounter{equation}{0}
\renewcommand{\theequation}{C.\arabic{equation}}
After each G\&R step, we update the constituent densities, rebuild the strain energy function with the current densities $\hat{\rho_i}$ (taken as median values across the myocardial wall), and load an incompressible myocardial tissue specimen with prescribed stretches. 
Herein, we enforce homogeneous deformations with the deformation gradient:
\begin{equation}
\ten{F} 
= \left[ 
  \begin{array}{lll}
  \lambda_{\rm{ff}} & \gamma_{\rm{fs}} & \gamma_{\rm{fn}} \\
  \gamma_{\rm{sf}} & \lambda_{\rm{ss}} & \gamma_{\rm{sn}} \\
  \gamma_{\rm{nf}} & \gamma_{\rm{ns}} & \lambda_{\rm{nn}} \\
  \end{array}
  \right],
\end{equation}
where the subscripts $\text{f, s, n}$ refers to the fiber, sheet, and sheet-normal directions. 
The resulting Cauchy stress is written as:
\begin{equation}
\ten{\sigma} 
= \left[ 
  \begin{array}{lll}
  \sigma_{\rm{ff}} & \sigma_{\rm{fs}} & \sigma_{\rm{fn}} \\
  \sigma_{\rm{sf}} & \sigma_{\rm{ss}} & \sigma_{\rm{sn}} \\
  \sigma_{\rm{nf}} & \sigma_{\rm{ns}} & \sigma_{\rm{nn}} \\
  \end{array}
  \right].
\end{equation}

We impose biaxial extension by incrementally increasing the fiber and sheet-normal stretches $\lambda_\text{ff}$ and $\lambda_\text{nn}$, while keeping all shear strains to zero. 
From the incompressibility condition, it follows that $\lambda_\text{ss} = \left( \lambda_\text{ff}\lambda_\text{nn} \right)^{-1}$ and for the zero-normal stress condition we have $\sigma_{\rm{ss}} = 0$. Finally, the biaxial test applied to the myocardial mixture gives:
\begin{equation}
\begin{aligned}
    \sigma_\text{ff} = &\  \sigma_\text{g,ff} + \sigma_\text{m,ff} + \sigma_\text{c,ff}\\
    =&\ 2 \hat{\rho}_\text{g} \frac{\partial\psi_\text{g}}{\partial I_\text{1}}(\lambda_\text{ff}^2 \ + \lambda_\text{ss}^2 ) +
     2 \hat{\rho}_\text{m} \frac{\partial\psi_\text{m}}{\partial I_\text{4mf}}\lambda_\text{ff}^2  \\
    +&\ 2 \hat{\rho}_\text{c} \frac{\partial\psi_\text{c}}{\partial I_\text{4cf}}\lambda_\text{ff}^2, 
\end{aligned}
\end{equation}
\begin{equation}
\begin{aligned}
    \sigma_\text{nn} = 
    &\ \sigma_\text{g,nn} + \sigma_\text{m,nn} + \sigma_\text{c,nn}\\
    =&\ 2 \hat{\rho}_\text{g} \frac{\partial\psi_\text{g}}{\partial I_\text{1}}(\lambda_\text{nn}^2 \ + \lambda_\text{ss}^2 ) + 2 \hat{\rho}_\text{c} \frac{\partial\psi_\text{c}}{\partial I_\text{4cf}}\lambda_\text{nn}^2.  
\end{aligned}
\end{equation}
Note that an isochoric–volumetric split was not introduced in the definition of the invariants, as incompressibility is enforced via a Lagrange multiplier.
\section*{Appendix D - Second Piola-Kirchhoff stress and elasticity tensor}
\label{sec:appendix_d}
\setcounter{equation}{0}
\renewcommand{\theequation}{D.\arabic{equation}}

Predicting the mechanical responses of the myocardial mixture requires deriving the second volumetric and isochoric Piola-Kirchhoff stresses as outlined in Eqs. (\ref{eq: volumetric stress}, \ref{eq: isochoric stress}), along with their derivatives with respect to the right Cauchy-Green tensor $\ten{C}^\text{e}$, the volumetric and isochoric elasticity tensors, see Eqs. (\ref{eq: volumetric elasticity tensor}, \ref{eq: isochoric elasticity tensor}).

The volumetric stress can be further expanded as:
\begin{equation}
\begin{aligned}
\ten{S}_\text{vol}
& = 2 \frac{\partial{\psi}_\text{vol}(J^\text{e})} {\partial\ten{C}^\text{e}},\\ 
& = 2 \frac{\partial{\psi}_\text{vol}(J^\text{e})}{\partial{J}^\text{e}} : \frac{\partial{J}^\text{e}}{\partial\ten{C}^\text{e}},\\
& = J^\text{e} p \ten{C}^{\text{e}-1}. 
\end{aligned}
\label{eq: volumetric stress derivation}
\end{equation}

Here, the hydrostatic pressure $p$ is computed from the constitutive relation in Eq.\eqref{eq: sedf volumetric}, 
\begin{equation}
p = \frac{d\psi_\text{vol}}{dJ^\text{e}} = \mu(J^\text{e} - 1).
\end{equation}

Taking the derivative of $\ten{S}_\text{vol}$ with respect to $\ten{C}^\text{e}$, we can further derive Eq.\eqref{eq: volumetric elasticity tensor} and retrieve:
\begin{equation}
\begin{aligned}
\mathbb{L}_\text{vol}
& = 2 \frac{\partial\ten{S}_\text{vol}}{\partial\ten{C}^\text{e}}\\
& = J^\text{e} \tilde{p}(\ten{C}^{\text{e}-1} \otimes \ten{C}^{\text{e}-1}) -2J^\text{e}p(\ten{C}^{\text{e}-1} \odot \ten{C}^{\text{e}-1}), 
\end{aligned}
\label{eq: volumetric elasticity tensor derivation}
\end{equation}
where,  for convenience, we introduced the operator $\odot$ such that, 
$-(\ten{C}^{\text{e}-1} \odot \ten{C}^{\text{e}-1})_{ijkl} = - \frac{1}{2}(C^{\text{e}-1}_{ik}C^{e-1}_{jl} + C^{\text{e}-1}_{il}C^{\text{e}-1}_{jk})$. 
$\{\CIRCLE \odot \Circle\}_{ijkl} = \frac{1}{2}(\{\CIRCLE\}_{ik} \{\Circle\}_{jl} + \{\CIRCLE\}_{il} \{\Circle\}_{jk})$ and $\{\CIRCLE \otimes\Circle\}_{ijkl} = \{\CIRCLE\}_{ij} \{\Circle\}_{kl}$. The scalar function $\tilde{p}$ is defined as $\tilde{p} = p + J(dp/dJ)$.

By integrating the microstructural constituents' description from Eq.\eqref{eq: strain energy constituent} into Eqs. (\ref{eq: strain energy}, \ref{eq: elastic stress}), we can expand the derivatives and retrieve for a generic constituent strain energy function $\psi_i$:  
\begin{equation}
\begin{aligned}
\ten{S}_{\text{iso},i} 
&= 2 \hat{\rho}_i \frac{\partial \Bar{\psi_i}(\Bar{\ten{C}})}{\partial \ten{C}^\text{e}}\\ 
&= 2 \hat{\rho}_i\frac{\partial \Bar{\psi_i}(\Bar{\ten{C}})}{\partial \Bar{\ten{C}}} : \frac{\partial \Bar{\ten{C}}}{\partial \ten{C}^\text{e}} \\
& = \Bar{\ten{S}}_i : J^{\text{e}-\frac{2}{3}} \left(\mathbb{I} - \frac{1}{3} \ten{C}^\text{e} \otimes \ten{C}^{\text{e}-1} \right) \\
& = J^{\text{e}-\frac{2}{3}} \mathbb{P} : \Bar{\ten{S}}_i \\
& = J^{\text{e}-\frac{2}{3}} \, \text{Dev}(\Bar{\ten{S}}_i),\\
\end{aligned}
\end{equation}
where the projection tensor is defined as $\mathbb{P} = \mathbb{I} - \frac{1}{3} \ten{C}^{\text{e}-1} \otimes \ten{C}^\text{e}$ and the deviatoric projection operator as $\text{Dev}(\CIRCLE) = (\CIRCLE) -(1/3) [(\CIRCLE) : \ten{C}^\text{e}]\ten{C}^{\text{e}-1}$ \citep{Holzapfel2006}. 
Therefore, for each constituent, it is essential to provide the fictitious stress, defined as $\ten{\Bar{S}}_i = 2\hat{\rho}_i{\partial \Bar{\psi}_i(\Bar{\ten{C}})}/{\partial\Bar{\ten{C}}}$.

\textbf{Ground matrix}\\
For the ground matrix, the second Piola-Kirchhoff stress takes the form: 
\begin{equation}
\ten{S}_{\text{iso,g}} 
= J^{\text{e}-\frac{2}{3}} \, \text{Dev}(\Bar{\ten{S}}_\text{g}) 
= 2 \hat{\rho}_\text{g} J^{\text{e}-\frac{2}{3}}  \gamma_{1\text{g}}\, \text{Dev}(\ten{I}),
\end{equation}
where we introduce $\gamma_{1\text{g}} = \partial \Bar{\psi}_\text{g} / \partial \Bar{I}_1.$ 

We further expand Eq.\eqref{eq: isochoric elasticity tensor} and find for the ground matrix constituent:
\begin{equation}
\begin{aligned}
\mathbb{L}_{\text{iso,g}} 
& = 2 \frac{\partial\ten{S}_{\text{iso,g}}}{\partial\ten{C}^\text{e}}\\
& = 4 \hat{\rho}_\text{g} J^{-\frac{4}{3}} \text{Dev}(\ten{I}) \otimes \text{Dev}(\ten{I})\\
& - \frac{4}{3} \hat{\rho}_\text{g} J^{-\frac{2}{3}} \gamma_{11\text{g}} \big(\ten{C}^{\text{e}-1} \otimes \text{Dev}(\ten{I})\\
&\qquad + \text{Dev}(\ten{I}) \otimes \ten{C}^{\text{e}-1} - \Bar{I}_1 \mathbb{\tilde P} \big),
\end{aligned}
\end{equation}
where we introduce $\gamma_{11\text{g}} = \partial \Bar{\psi}_\text{g}^2 / \partial \Bar{I}_{1}^2,$ 
and the modified projection tensor $\tilde{\mathbb{P}} = \ten{C}^{-1} \odot \ten{C}^{-1} - \frac{1}{3} \ten{C}^{-1} \otimes \ten{C}^{-1}.$

\textbf{Collagen fibers}\\
Collagen fibers isochoric contribution to the elastic stress includes the tension-related terms and the shear term, see Eq.\eqref{eq:collagen_constitutive}. We can, thus, compute it as:
\begin{equation}
\begin{aligned}
\ten{S}_{\text{iso,c}} 
& = J^{\text{e}-\frac{2}{3}}\text{Dev}(\Bar{\ten{S}}_\text{c})\\
& = 2 \hat{\rho}_\text{c} J^{\text{e}-\frac{2}{3}} \sum_{i \in \{\text{f,s,n}\}} \gamma_{4\text{c}i}\, \text{Dev} \left(\ten{A}_{0,i} \right)\\
& + \hat{\rho}_\text{c} J^{\text{e}-\frac{2}{3}}\gamma_{8\text{c}}\, \text{Dev} \left(\ten{S}_{0} \right),
\end{aligned}
\end{equation}
where we introduce $\gamma_{4\text{c}i} = \partial \Bar{\psi}_{\text{c}i} / \partial \Bar{I}_{4\text{c}i}$ for the tensile strain energy terms
and $\gamma_{8\text{c}} = \partial \Bar{\psi}_\text{cfs} / \partial \Bar{I}_\text{8cfs}$ for the shear one as defined hereunder:

\begin{equation}
\Bar{\psi}_{\text{c}i}(\Bar{I}_{4\text{c}i})
= \frac{a_{\text{c}i}}{2b_{\text{c}i}}
\left(
\exp\left(b_{\text{c}i}\langle \Bar{I}_{4\text{c}i}-1 \rangle^2\right)-1
\right),
\end{equation}
\begin{equation}
\Bar{\psi}_{\text{cfs}}(\Bar{I}_{8\text{cfs}})
= \frac{a_{\text{cfs}}}{2b_{\text{cfs}}}
\left(
\exp\left(b_{\text{cfs}}(\Bar{I}_{8\text{cfs}})^2\right)-1
\right).
\end{equation}

The structural direction tensor $\ten{A}_{0,i} = \partial \Bar{I}_{4\text{c}i} / \partial\Bar{\ten{C}}$ in the undeformed configuration is specified in each direction, i.e., $\ten{A}_{0,\text{f}} = \vec{f}_0 \otimes \vec{f}_0$, $\ten{A}_{0,\text{s}} = \vec{s}_0 \otimes \vec{s}_0$, and $\ten{A}_{0,\text{n}} = \vec{n}_0 \otimes \vec{n}_0$. 
Finally, we introduce a structural tensor in the \textit{fs}-plane, $\ten{S}_{0} = \partial \Bar{I}_\text{8\text{c}fs} / \partial\Bar{\ten{C}} = (\vec{f}_0 \otimes \vec{s}_0 + \vec{s}_0 \otimes \vec{f}_0)/2$.

The isochoric contribution of the elasticity tensor related to collagen fibers can be expressed as:
\begin{equation}
\begin{aligned}
\mathbb{L}_{\text{iso,c}}
& = 2 \frac{\partial\ten{S}_{\text{iso,c}}}{\partial\ten{C}^\text{e}}\\
& = 4 \hat{\rho}_\text{c} J^{-\frac{4}{3}} \sum_{i \in \{\text{f,s,n}\}} \gamma_{44\text{c}i}\, \text{Dev} (\ten{A}_{0,i}) \otimes \text{Dev} (\ten{A}_{0,i})\\
     & - \frac{4}{3} \hat{\rho}_\text{c} J^{-\frac{2}{3}} \sum_{i \in \{\text{f,s,n}\}} \gamma_{4\text{c}i} \big( \ten{C}^{\text{e}-1} \otimes \text{Dev}(\ten{A}_{0,i})\\
     &\qquad + \text{Dev}(\ten{A}_{0,i}) \otimes \ten{C}^{\text{e}-1}  - \Bar{I}_{4\text{c}i} \mathbb{\tilde P} \big) \\
     & + 4 \hat{\rho}_\text{c} J^{-\frac{4}{3}} \gamma_{88\text{c}}\, \text{Dev}(\ten{S}_{0}) \otimes \text{Dev}(\ten{S}_{0})\\
     & - \frac{4}{3} \hat{\rho}_\text{c} J^{-\frac{2}{3}} \gamma_{8\text{c}} \big( \ten{C}^{\text{e}-1} \otimes \text{Dev} (\ten{S}_{0})\\ 
     &\qquad + \text{Dev}(\ten{S}_{0}) \otimes \ten{C}^{\text{e}-1}  - \Bar{I}_{8} \mathbb{\tilde P} \big),
\end{aligned}
\label{eq:isochoric_elasticity_tensor_collagen_fibers}
\end{equation}
where we introduce the second order derivatives $\gamma_{44\text{c}i} = \partial \Bar{\psi}_{\text{c}i}^2/\partial \Bar{I}_{4\text{c}i}^2$ 
and $\gamma_{88\text{c}} = \partial \Bar{\psi}_\text{cfs}^2 / \partial \Bar{I}_\text{8cfs}^2.$

\textbf{Myocytes}\\
Based on the quasi-one-dimensional anisotropic strain energy contribution of myocytes in Eq.~\eqref{eq: myocytes sedf}, the isochoric second Piola–Kirchhoff stress tensor is given by:
\begin{equation}
\begin{aligned}
\ten{S}_{\text{iso,m}} 
& = J^{\text{e}-\frac{2}{3}} \, \text{Dev}(\Bar{\ten{S}}_\text{m}) \\
& = \hat{\rho}_\text{m} J^{\text{e}-\frac{2}{3}} \gamma_\text{4m} \text{Dev}(\ten{M}_0),
\end{aligned}
\end{equation}
where we introduce $\gamma_\text{4m} = \partial \Bar{\psi}_\text{m}/\partial \Bar{I}_\text{4m}$

and the structural tensor of the myocyte in the undeformed configuration, $\ten{M}_{0} = \vec{f}_0 \otimes \vec{f}_0$. 
Note that the latter coincides with the direction of the collagen fiber aligned with the longitudinal axis of the myofiber.

We can finally write the elasticity tensor associated with myocytes as:
\begin{equation}
\begin{aligned}
\mathbb{L}_{\text{iso,m}}  
& = 2 \frac{\partial\ten{S}_{\text{iso,m}}}{\partial\ten{C}^\text{e}}\\
& = 4 \hat{\rho}_\text{m} J^{-\frac{4}{3}} \gamma_\text{44m} \text{Dev}(\ten{M}_0) \otimes \text{Dev}(\ten{M}_0)\\
     &- \frac{4}{3} \hat{\rho}_\text{m} J^{-\frac{2}{3}} \gamma_\text{4m} \big( \ten{C}^{\text{e}-1} \otimes \text{Dev}(\ten{M}_0)\\
     &\qquad + \text{Dev}(\ten{M}_0) \otimes \ten{C}^{\text{e}-1}  - \Bar{I}_\text{4m} \mathbb{\tilde P} \big),\\
\end{aligned}
\label{eq: isochoric elasticity tensor myocytes}
\end{equation}
where we introduce $\gamma_\text{44m} = \partial \Bar{\psi}_\text{m}^2/\partial \Bar{I}_\text{4m}^2$. 
The collagen and myocyte strain energy contributions use modified isochoric invariants for anisotropic fiber terms. 
This formulation is consistent with the constitutive model selected in this study and is common in nearly incompressible soft-tissue formulations. 
However, applying the deviatoric projection to anisotropic fiber terms can induce undesired volumetric-stress coupling under large deformations \citep{Gultekin2019OnMaterials}. 
As the present simulations involve moderate deformation ranges, this limitation is not expected to affect the qualitative trends reported here. 
Nevertheless, future extensions could alternatively apply the isochoric split only to the isotropic matrix contribution, while formulating the anisotropic fiber terms in terms of the corresponding full invariants \citep{Sansour2008OnAnisotropy,Helfenstein2010OnMaterials}.
\newpage

 \bibliographystyle{elsarticle-num}
 \bibliography{library}

@article{Conrad1995,
author = {Chester H. Conrad  and Wesley W. Brooks  and John A. Hayes  and Subha Sen  and Kathleen G. Robinson  and Oscar H. L. Bing },
title = {Myocardial Fibrosis and Stiffness With Hypertrophy and Heart Failure in the Spontaneously Hypertensive Rat},
journal = {Circulation},
volume = {91},
number = {1},
pages = {161-170},
year = {1995},
doi = {10.1161/01.CIR.91.1.161},
}

@article{Wilson2022,
   author = {Alexander J. Wilson and Gregory B. Sands and Ian J. LeGrice and Alistair A. Young and Daniel B. Ennis},
   doi = {10.1152/AJPHEART.00059.2022},
   journal = {American Journal of Physiology - Heart and Circulatory Physiology},
   publisher = {American Physiological Society},
   title = {Myocardial Mesostructure and Mesofunction},
   volume = {323},
   year = {2022}
}

@article{Sommer2015,
    author = {Sommer, Gerhard and Schriefl, Andreas J. and Andrä, Michaela and Sacherer, Michael and Viertler, Christian and Wolinski, Heimo and Holzapfel, Gerhard A.},
   doi = {10.1016/j.actbio.2015.06.031},
   journal = {Acta Biomaterialia},
   month = {9},
   pages = {172-192},
   publisher = {Elsevier Ltd},
   title = {Biomechanical properties and microstructure of human ventricular myocardium},
   volume = {24},
   year = {2015}
}

@article{Gebauer2024b,
   author = {Amadeus M. Gebauer and Martin R. Pfaller and Jason M. Szafron and Wolfgang A. Wall},
   doi = {10.1002/cnm.3869},
   issn = {20407947},
   issue = {11},
   journal = {International Journal for Numerical Methods in Biomedical Engineering},
   month = {11},
   pmid = {39300801},
   publisher = {John Wiley and Sons Inc},
   title = {Adaptive integration of history variables in constrained mixture models for organ-scale growth and remodeling},
   volume = {40},
   year = {2024}
}

@article{Witzenburg2017,
   author = {Colleen M. Witzenburg and Jeffrey W. Holmes},
   doi = {10.1007/s10659-017-9631-8},
   issn = {15732681},
   issue = {1-2},
   journal = {Journal of Elasticity},
   month = {12},
   pages = {257-281},
   publisher = {Springer Netherlands},
   title = {A Comparison of Phenomenologic Growth Laws for Myocardial Hypertrophy},
   volume = {129},
   year = {2017}
}

@article{Guccione1993,
    author = {Guccione, J. M. and Waldman, L. K. and McCulloch, A. D.},
    title = {Mechanics of Active Contraction in Cardiac Muscle: Part II—Cylindrical Models of the Systolic Left Ventricle},
    journal = {Journal of Biomechanical Engineering},
    volume = {115},
    number = {1},
    pages = {82-90},
    year = {1993},
    month = {02},
    issn = {0148-0731},
    doi = {10.1115/1.2895474},
}

@article{Gktepe2010,
   author = {Serdar Göktepe and Oscar John Abilez and Ellen Kuhl},
   doi = {10.1016/j.jmps.2010.07.003},
   issn = {00225096},
   issue = {10},
   journal = {Journal of the Mechanics and Physics of Solids},
   month = {10},
   pages = {1661-1680},
   title = {A generic approach towards finite growth with examples of athlete's heart, cardiac dilation, and cardiac wall thickening},
   volume = {58},
   year = {2010}
}

@misc{Wang2015,
   author = {V. Y. Wang and P. M.F. Nielsen and M. P. Nash},
   doi = {10.1146/annurev-bioeng-071114-040609},
   issn = {15454274},
   journal = {Annual Review of Biomedical Engineering},
   month = {12},
   pages = {351-383},
   pmid = {26643023},
   publisher = {Annual Reviews Inc.},
   title = {Image-Based Predictive Modeling of Heart Mechanics},
   volume = {17},
   year = {2015}
}

@article{Sacks2003,
   author = {Michael S. Sacks},
   doi = {10.1115/1.1544508},
   issn = {01480731},
   issue = {2},
   journal = {Journal of Biomechanical Engineering},
   month = {4},
   pages = {280-287},
   pmid = {12751291},
   title = {Incorporation of experimentally-derived fiber orientation into a structural constitutive model for planar collagenous tissues},
   volume = {125},
   year = {2003}
}

@article{Nevo1989,
    author = {Nevo, E. and Lanir, Y.},
    title = {Structural Finite Deformation Model of the Left Ventricle During Diastole and Systole},
    journal = {Journal of Biomechanical Engineering},
    volume = {111},
    number = {4},
    pages = {342-349},
    year = {1989},
    month = {11},
    issn = {0148-0731},
    doi = {10.1115/1.3168389},
}

@misc{Wisdom2015,
   author = {Katrina M. Wisdom and Scott L. Delp and Ellen Kuhl},
   doi = {10.1007/s10237-014-0607-3},
   issn = {16177940},
   issue = {2},
   journal = {Biomechanics and Modeling in Mechanobiology},
   month = {4},
   pages = {195-215},
   pmid = {25199941},
   publisher = {Springer Verlag},
   title = {Use it or lose it: multiscale skeletal muscle adaptation to mechanical stimuli},
   volume = {14},
   year = {2015},
}

@misc{Emery1997,
   author = {Jeff L Emery and Jeffrey H Omens},
   title = {Mechanical regulation of myocardial growth during volume-overload hypertrophy in the rat},
   year = {1997},
}

@misc{Opie2006,
   author = {Lionel H Opie and Patrick J Commerford and Bernard J Gersh and Marc A Pfeffer},
   journal = {www.thelancet.com},
   title = {Series Controversies in Cardiology 4 Controversies in ventricular remodelling},
   volume = {367},
   year = {2006},
}

@article{Zeng2017,
   author = {Decai Zeng and Hui Chen and Chun Lan Jiang and Ji Wu},
   doi = {10.1097/MD.0000000000007968},
   issn = {15365964},
   issue = {36},
   journal = {Medicine (United States)},
   month = {9},
   pmid = {28885350},
   publisher = {Lippincott Williams and Wilkins},
   title = {Usefulness of three-dimensional spherical index to assess different types of left ventricular remodeling},
   volume = {96},
   year = {2017},
}

@article{Kerckhoffs2012,
   author = {Roy C.P. Kerckhoffs and Jeffrey H. Omens and Andrew D. McCulloch},
   doi = {10.1016/j.mechrescom.2011.11.004},
   issn = {00936413},
   journal = {Mechanics Research Communications},
   month = {6},
   pages = {40-50},
   pmid = {22639476},
   title = {A single strain-based growth law predicts concentric and eccentric cardiac growth during pressure and volume overload},
   volume = {42},
   year = {2012}
}

@article{Li2003,
author = {Li, Sherry and Demmel, James},
year = {2003},
month = {06},
pages = {110-140},
title = {SuperLUDIST: A scalable distributed-memory sparse direct solver for unsymmetric linear systems},
volume = {29},
journal = {ACM Trans. Math. Softw.},
doi = {10.1145/779359.779361}
}

@article{Wang2013,
   author = {H. M. Wang and H. Gao and X. Y. Luo and C. Berry and B. E. Griffith and R. W. Ogden and T. J. Wang},
   doi = {10.1002/cnm.2497},
   issn = {20407939},
   issue = {1},
   journal = {International Journal for Numerical Methods in Biomedical Engineering},
   month = {1},
   pages = {83-103},
   pmid = {23293070},
   title = {Structure-based finite strain modelling of the human left ventricle in diastole},
   volume = {29},
   year = {2013}
}

@article{Lombaert2012,
   author = {Herve Lombaert and Jean Marc Peyrat and Pierre Croisille and Stanislas Rapacchi and Laurent Fanton and Farida Cheriet and Patrick Clarysse and Isabelle Magnin and Hervé Delingette and Nicholas Ayache},
   doi = {10.1109/TMI.2012.2192743},
   issn = {02780062},
   issue = {7},
   journal = {IEEE Transactions on Medical Imaging},
   pages = {1436-1447},
   pmid = {22481815},
   title = {Human atlas of the cardiac fiber architecture: Study on a healthy population},
   volume = {31},
   year = {2012}
}

@article{Jia2019,
   author = {Zheng Jia and Thao D. Nguyen},
   doi = {10.1016/j.jmbbm.2019.06.004},
   issn = {18780180},
   journal = {Journal of the Mechanical Behavior of Biomedical Materials},
   month = {10},
   pages = {96-107},
   pmid = {31212199},
   publisher = {Elsevier Ltd},
   title = {A micromechanical model for the growth of collagenous tissues under mechanics-mediated collagen deposition and degradation},
   volume = {98},
   year = {2019}
}

@misc{Eriksson2013,
   author = {Thomas S.E. Eriksson and Anton J. Prassl and Gernot Plank and Gerhard A. Holzapfel},
   doi = {10.1002/cnm.2575},
   issn = {20407939},
   issue = {11},
   journal = {International Journal for Numerical Methods in Biomedical Engineering},
   month = {11},
   pages = {1267-1284},
   pmid = {23868817},
   title = {Modeling the dispersion in electromechanically coupled myocardium},
   volume = {29},
   year = {2013}
}

@misc{Holzapfel2009,
   author = {Gerhard A. Holzapfel and Ray W. Ogden},
   doi = {10.1098/rsta.2009.0091},
   issn = {1364503X},
   issue = {1902},
   journal = {Philosophical Transactions of the Royal Society A: Mathematical, Physical and Engineering Sciences},
   month = {9},
   pages = {3445-3475},
   pmid = {19657007},
   publisher = {Royal Society},
   title = {Constitutive modelling of passive myocardium: A structurally based framework for material characterization},
   volume = {367},
   year = {2009}
}

@misc{Liu2025,
   author = {Taiwei Liu and Fuyou Liang},
   doi = {10.1016/j.mechmat.2025.105273},
   issn = {01676636},
   journal = {Mechanics of Materials},
   month = {4},
   publisher = {Elsevier B.V.},
   title = {A microstructure-based finite element model of the human left ventricle for simulating the trans-scale myocardial mechanical behaviors},
   volume = {203},
   year = {2025}
}

@misc{McEvoy2018,
   author = {Eoin McEvoy and Gerhard A. Holzapfel and Patrick McGarry},
   doi = {10.1115/1.4039947},
   issn = {15288951},
   issue = {8},
   journal = {Journal of Biomechanical Engineering},
   month = {8},
   pmid = {30003247},
   publisher = {American Society of Mechanical Engineers (ASME)},
   title = {Compressibility and Anisotropy of the Ventricular Myocardium: Experimental Analysis and Microstructural Modeling},
   volume = {140},
   year = {2018}
}

@misc{Xi2019,
   author = {Ce Xi and Ghassan S. Kassab and Lik Chuan Lee},
   doi = {10.1016/j.actbio.2019.04.016},
   issn = {18787568},
   journal = {Acta Biomaterialia},
   month = {5},
   pages = {241-253},
   pmid = {30980939},
   publisher = {Acta Materialia Inc},
   title = {Microstructure-based finite element model of left ventricle passive inflation},
   volume = {90},
   year = {2019}
}

@misc{Jin2011,
   author = {Yu Fang Jin and Hai Chao Han and Jamie Berger and Qiuxia Dai and Merry L. Lindsey},
   doi = {10.1186/1752-0509-5-60},
   issn = {17520509},
   journal = {BMC Systems Biology},
   month = {5},
   pmid = {21545710},
   title = {Combining experimental and mathematical modeling to reveal mechanisms of macrophage-dependent left ventricular remodeling},
   volume = {5},
   year = {2011},
}

@misc{Sesa2023,
   author = {Mahmoud Sesa and Hagen Holthusen and Lukas Lamm and Christian Böhm and Tim Brepols and Stefan Jockenhövel and Stefanie Reese},
   doi = {10.1016/j.compbiomed.2023.107623},
   issn = {18790534},
   journal = {Computers in Biology and Medicine},
   month = {12},
   pmid = {37922603},
   publisher = {Elsevier Ltd},
   title = {Mechanical modeling of the maturation process for tissue-engineered implants: Application to biohybrid heart valves},
   volume = {167},
   year = {2023},
}

@article{Sesa2025,
   author = {Mahmoud Sesa and Hagen Holthusen and Christian Böhm and Stefan Jockenhövel and Stefanie Reese and Kevin Linka},
   month = {3},
   title = {A Comprehensive Framework for Predictive Computational Modeling of Growth and Remodeling in Tissue-Engineered Cardiovascular Implants},
   url = {http://arxiv.org/abs/2503.17151},
   journal = {Arxiv},
   year = {2025}
}

@misc{Saez2013,
   author = {Pablo Sáez and Estefanía Peña and Miguel Ángel Martínez and Ellen Kuhl},
   doi = {10.1007/s00285-012-0613-y},
   issn = {03036812},
   issue = {6-7},
   journal = {Journal of Mathematical Biology},
   month = {12},
   pages = {1765-1793},
   pmid = {23129392},
   title = {Mathematical modeling of collagen turnover in biological tissue},
   volume = {67},
   year = {2013},
}

@misc{Grytsan2017,
   author = {Andrii Grytsan and Thomas S.E. Eriksson and Paul N. Watton and T. Christian Gasser},
   doi = {10.3390/ma10090994},
   issn = {19961944},
   issue = {9},
   journal = {Materials},
   month = {8},
   publisher = {MDPI AG},
   title = {Growth description for VesselWall adaptation: A Thick-Walled mixture model of abdominal aortic aneurysm evolution},
   volume = {10},
   year = {2017},
}

@misc{Eriksson2014,
   author = {T. S.E. Eriksson and P. N. Watton and X. Y. Luo and Y. Ventikos},
   doi = {10.1016/j.jmps.2014.09.003},
   issn = {00225096},
   journal = {Journal of the Mechanics and Physics of Solids},
   month = {12},
   pages = {134-150},
   publisher = {Elsevier Ltd},
   title = {Modelling volumetric growth in a thick walled fibre reinforced artery},
   volume = {73},
   year = {2014},
}

@book{Holzapfel2006,
   author = {Gerhard A. Holzapfel},
   isbn = {0471823198},
   pages = {455},
   publisher = {John Wiley \& Sons},
   title = {Nonlinear solid mechanics : a continuum approach for engineering},
   year = {2006},
}

@misc{Goktepe2011,
   author = {S. Göktepe and S. N.S. Acharya and J. Wong and E. Kuhl},
   doi = {10.1002/cnm.1402},
   issn = {20407939},
   issue = {1},
   journal = {International Journal for Numerical Methods in Biomedical Engineering},
   month = {1},
   pages = {1-12},
   title = {Computational modeling of passive myocardium},
   volume = {27},
   year = {2011},
}

@book{Goriely2017,
author = {Goriely, Alain},
year = {2017},
month = {01},
pages = {},
title = {The Mathematics and Mechanics of Biological Growth},
volume = {45},
isbn = {978-0-387-87709-9},
doi = {10.1007/978-0-387-87710-5},
publisher = {Springer New York, NY}
}

@misc{Humphrey2002,
   author = {J D Humphrey and K R Rajagopal},
   issue = {3},
   pages = {407-430},
   title = {A CONSTRAINED MIXTURE MODEL FOR GROWTH AND REMODELING OF SOFT TISSUES},
   volume = {16},
   year = {2002},
}

@misc{Gebauer2023,
   author = {Amadeus M. Gebauer and Martin R. Pfaller and Fabian A. Braeu and Christian J. Cyron and Wolfgang A. Wall},
   doi = {10.1007/s10237-023-01747-w},
   issn = {16177940},
   issue = {6},
   journal = {Biomechanics and Modeling in Mechanobiology},
   month = {12},
   pages = {1983-2002},
   pmid = {37482576},
   publisher = {Springer Science and Business Media Deutschland GmbH},
   title = {A homogenized constrained mixture model of cardiac growth and remodeling: analyzing mechanobiological stability and reversal},
   volume = {22},
   year = {2023},
}

@misc{Sharifi2021,
   author = {Hossein Sharifi and Charles K Mann and Alexus L Rockward and Mohammad Mehri and Joy Mojumder and Lik-Chuan Lee and Kenneth S Campbell and Jonathan F Wenk},
   doi = {10.1007/s12551-021-00826-5/Published},
   issue = {5},
   journal = {Biophysical Reviews},
   pages = {729-746},
   title = {Multiscale simulations of left ventricular growth and remodeling},
   volume = {13},
   year = {2021},
}

@misc{Spinale2000,
   author = {Francis G Spinale and Mytsi L Coker and Brian R Bond and James L Zellner},
   journal = {Cardiovascular Research},
   pages = {225-238},
   title = {Myocardial matrix degradation and metalloproteinase activation in the failing heart: a potential therapeutic target},
   volume = {46},
   year = {2000},
    doi = {10.1016/s0008-6363(99)00431-9}
}

@misc{Cleutjens1996,
   author = {Jack P M Cleutjens},
   issue = {816},
   journal = {Cardiovascular Research},
   pages = {816-821},
   title = {Cardiovascular Mystery Series The role of matrix metalloproteinases in heart disease},
   volume = {32},
   year = {1996},
}

@misc{Schmid2012,
   author = {H. Schmid and L. Pauli and A. Paulus and E. Kuhl and M. Itskov},
   doi = {10.1080/10255842.2010.548325},
   issn = {10255842},
   issue = {5},
   journal = {Computer Methods in Biomechanics and Biomedical Engineering},
   month = {5},
   pages = {547-561},
   pmid = {21347909},
   title = {Consistent formulation of the growth process at the kinematic and constitutive level for soft tissues composed of multiple constituents},
   volume = {15},
   year = {2012},
}

@misc{Laubrie2022,
   author = {Joan D. Laubrie and S. Jamaleddin Mousavi and Stéphane Avril},
   doi = {10.1007/s10237-021-01544-3},
   issn = {16177940},
   issue = {2},
   journal = {Biomechanics and Modeling in Mechanobiology},
   month = {4},
   pages = {455-469},
   pmid = {35067825},
   publisher = {Springer Science and Business Media Deutschland GmbH},
   title = {About prestretch in homogenized constrained mixture models simulating growth and remodeling in patient-specific aortic geometries},
   volume = {21},
   year = {2022},
}

@misc{Guan2023,
   author = {Debao Guan and Xin Zhuan and Xiaoyu Luo and Hao Gao},
   doi = {10.1016/j.actbio.2023.05.022},
   issn = {18787568},
   journal = {Acta Biomaterialia},
   month = {8},
   pages = {375-399},
   pmid = {37201740},
   publisher = {Acta Materialia Inc},
   title = {An updated Lagrangian constrained mixture model of pathological cardiac growth and remodelling},
   volume = {166},
   year = {2023},
}

@misc{Kroon2009,
   author = {Wilco Kroon and Tammo Delhaas and Theo Arts and Peter Bovendeerd},
   doi = {10.1007/s10237-008-0136-z},
   issn = {16822978},
   issue = {4},
   journal = {Biomechanics and Modeling in Mechanobiology},
   pages = {301-309},
   pmid = {18758835},
   publisher = {Springer Verlag},
   title = {Computational modeling of volumetric soft tissue growth: Application to the cardiac left ventricle},
   volume = {8},
   year = {2009},
}

@misc{Genet2016,
   author = {M. Genet and L. C. Lee and B. Baillargeon and J. M. Guccione and E. Kuhl},
   doi = {10.1007/s10439-015-1351-2},
   issn = {15739686},
   issue = {1},
   journal = {Annals of Biomedical Engineering},
   month = {1},
   pages = {112-127},
   pmid = {26043672},
   publisher = {Springer New York LLC},
   title = {Modeling Pathologies of Diastolic and Systolic Heart Failure},
   volume = {44},
   year = {2016},
}

@misc{Rodriguez1994,
   author = {Edward K Rodriguez and Anne Hoger and Andrew D Mcculloch},
   issue = {4},
   journal = {Pergamon .I. Biomchanics},
   pages = {455-467},
   title = {STRESS-DEPENDENT FINITE GROWTH IN SOFT ELASTIC TISSUES},
   volume = {21},
   year = {1994},
}

@misc{Lee2016,
   author = {L. C. Lee and G. S. Kassab and J. M. Guccione},
   doi = {10.1002/wsbm.1330},
   issn = {1939005X},
   issue = {3},
   journal = {Wiley Interdisciplinary Reviews: Systems Biology and Medicine},
   month = {5},
   pages = {211-226},
   pmid = {26952285},
   publisher = {Wiley-Blackwell},
   title = {Mathematical modeling of cardiac growth and remodeling},
   volume = {8},
   year = {2016},
}

@misc{Pathak2001,
   author = {Manas Pathak and Sagartirtha Sarkar and Elangovan Vellaichamy and Subha Sen},
   title = {Role of Myocytes in Myocardial Collagen Production},
   url = {http://www.hypertensionaha.org},
   year = {2001},
}

@misc{Melenovsky2013,
   author = {Vojtech Melenovsky},
   doi = {10.1007/978-1-4614-5203-4_9},
   isbn = {9781461452034},
   journal = {Cardiac Adaptations: Molecular Mechanisms},
   month = {1},
   pages = {167-199},
   publisher = {Springer New York},
   title = {Cardiac adaptation to volume overload},
   volume = {4},
   year = {2013},
}

@misc{Savarese2022,
   author = {Gianluigi Savarese and Peter Moritz Becher and Lars H. Lund and Petar Seferovic and Giuseppe M.C. Rosano and Andrew J.S. Coats},
   doi = {10.1093/cvr/cvac013},
   issn = {17553245},
   issue = {17},
   journal = {Cardiovascular Research},
   month = {12},
   pages = {3272-3287},
   pmid = {35150240},
   publisher = {Oxford University Press},
   title = {Global burden of heart failure: a comprehensive and updated review of epidemiology},
   volume = {118},
   year = {2022},
}

@misc{Weber1989,
   author = {Karl T. Weber},
   doi = {10.1016/0735-1097(89)90360-4},
   issn = {07351097},
   issue = {7},
   journal = {Journal of the American College of Cardiology},
   pages = {1637-1652},
   pmid = {2656824},
   title = {Cardiac interstitium in health and disease: The fibrillar collagen network},
   volume = {13},
   year = {1989},
}

@misc{Corporan2018,
   author = {Daniella Corporan and Daisuke Onohara and Roberto Hernandez-Merlo and Alicja Sielicka and X Muralidhar Padala},
   doi = {10.1152/ajpheart.00099.2018.-Mitral},
   journal = {Am J Physiol Heart Circ Physiol},
   pages = {1269-1278},
   title = {Temporal changes in myocardial collagen, matrix metalloproteinases, and their tissue inhibitors in the left ventricular myocardium in experimental chronic mitral regurgitation in rodents},
   volume = {315},
   year = {2018},
}

@misc{Ryan2007,
   author = {Thomas D. Ryan and Emily C. Rothstein and Inmaculada Aban and Jose A. Tallaj and Ahsan Husain and Pamela A. Lucchesi and Louis J. Dell'Italia},
   doi = {10.1016/j.jacc.2006.06.083},
   issn = {07351097},
   issue = {7},
   journal = {Journal of the American College of Cardiology},
   month = {2},
   pages = {811-821},
   pmid = {17306712},
   title = {Left Ventricular Eccentric Remodeling and Matrix Loss Are Mediated by Bradykinin and Precede Cardiomyocyte Elongation in Rats With Volume Overload},
   volume = {49},
   year = {2007},
}

@misc{Bonow1992,
    author = { Robert O. Bonow and James E. Udelson},
    title = {Left Ventricular Diastolic Dysfunction as a Cause of Congestive Heart Failure},
    journal = {Annals of Internal Medicine},
    volume = {117},
    number = {6},
    pages = {502-510},
    year = {1992},
    doi = {10.7326/0003-4819-117-6-502},
}

@misc{Hutchinson2010,
   author = {Kirk R. Hutchinson and James A. Stewart and Pamela A. Lucchesi},
   doi = {10.1016/j.yjmcc.2009.06.001},
   issn = {00222828},
   issue = {3},
   journal = {Journal of Molecular and Cellular Cardiology},
   month = {3},
   pages = {564-569},
   pmid = {19524591},
   title = {Extracellular matrix remodeling during the progression of volume overload-induced heart failure},
   volume = {48},
   year = {2010},
}

@misc{Grossman1980,
   author = {William Grossman},
   title = {Cardiac Hypertrophy: Useful Adaptation or Pathologic Process?},
    issue = {69},
    doi = {/10.1016/0002-9343(80)90471-4},
   journal = {The American Journal of Medicine},
   year = {1980},
}

@misc{Nakamura2018,
   author = {Michinari Nakamura and Junichi Sadoshima},
   doi = {10.1038/s41569-018-0007-y},
   issn = {17595010},
   issue = {7},
   journal = {Nature Reviews Cardiology},
   month = {7},
   pages = {387-407},
   pmid = {29674714},
   publisher = {Nature Publishing Group},
   title = {Mechanisms of physiological and pathological cardiac hypertrophy},
   volume = {15},
   year = {2018},
}

@Article{Peirlinck2017,
  author    = {Peirlinck, Mathias and Debusschere, Nic and Iannaccone, Francesco and Siersema, Peter D. and Verhegghe, Benedict and Segers, Patrick and De Beule, Matthieu},
  journal   = {Biomechanics and Modeling in Mechanobiology},
  title     = {An in silico biomechanical analysis of the stent–esophagus interaction},
  year      = {2017},
  issn      = {1617-7940},
  month     = aug,
  number    = {1},
  pages     = {111--131},
  volume    = {17},
  doi       = {10.1007/s10237-017-0948-9},
  publisher = {Springer Science and Business Media LLC},
}

@Article{Martonova2024,
  author    = {Martonová, Denisa and Peirlinck, Mathias and Linka, Kevin and Holzapfel, Gerhard A. and Leyendecker, Sigrid and Kuhl, Ellen},
  journal   = {Computer Methods in Applied Mechanics and Engineering},
  title     = {Automated model discovery for human cardiac tissue: Discovering the best model and parameters},
  year      = {2024},
  issn      = {0045-7825},
  month     = aug,
  pages     = {117078},
  volume    = {428},
  doi       = {10.1016/j.cma.2024.117078},
  publisher = {Elsevier BV},
}

@Article{Peirlinck2018b,
  author    = {Peirlinck, Mathias and Sack, Kevin L. and De Backer, Pieter and Morais, Pedro and Segers, Patrick and Franz, Thomas and De Beule, Matthieu},
  journal   = {International Journal for Numerical Methods in Biomedical Engineering},
  title     = {Kinematic boundary conditions substantially impact in silico ventricular function},
  year      = {2018},
  issn      = {2040-7947},
  month     = oct,
  number    = {1},
  volume    = {35},
  doi       = {10.1002/cnm.3151},
  publisher = {Wiley},
}

@Article{Peirlinck2018a,
  author    = {Peirlinck, Mathias and De Beule, Matthieu and Segers, Patrick and Rebelo, Nuno},
  journal   = {Journal of the Mechanical Behavior of Biomedical Materials},
  title     = {A modular inverse elastostatics approach to resolve the pressure-induced stress state for in vivo imaging based cardiovascular modeling},
  year      = {2018},
  issn      = {1751-6161},
  month     = sep,
  pages     = {124--133},
  volume    = {85},
  doi       = {10.1016/j.jmbbm.2018.05.032},
  publisher = {Elsevier BV},
}

@Article{Peirlinck2019,
  author    = {Peirlinck, M. and Sahli Costabal, F. and Sack, K. L. and Choy, J. S. and Kassab, G. S. and Guccione, J. M. and De Beule, M. and Segers, P. and Kuhl, E.},
  journal   = {Biomechanics and Modeling in Mechanobiology},
  title     = {Using machine learning to characterize heart failure across the scales},
  year      = {2019},
  issn      = {1617-7940},
  month     = jun,
  number    = {6},
  pages     = {1987--2001},
  volume    = {18},
  doi       = {10.1007/s10237-019-01190-w},
  publisher = {Springer Science and Business Media LLC},
}

@Misc{Cardona2025,
  author    = {Cardona, Sara and Peirlinck, Mathias and Fereidoonnezhad, Behrooz},
  title     = {TopoGEN: topology-driven microstructure generation for in silico modeling of fiber network mechanics},
  year      = {2025},
  copyright = {Creative Commons Attribution Non Commercial Share Alike 4.0 International},
  doi       = {10.48550/ARXIV.2503.19832},
  publisher = {arXiv},
}

@article{Wang2016Image-drivenFibrosis,
    title = {{Image-driven constitutive modeling of myocardial fibrosis}},
    year = {2016},
    journal = {International Journal for Computational Methods in Engineering Science and Mechanics},
    author = {Wang, Vicky Y. and Niestrawska, Justyna A. and Wilson, Alexander J. and Sands, Gregory B. and Young, Alistair A. and LeGrice, Ian J. and Nash, Martyn P.},
    number = {3},
    month = {5},
    pages = {211--221},
    volume = {17},
    publisher = {Bellwether Publishing, Ltd.},
    doi = {10.1080/15502287.2015.1082675},
    issn = {15502295}
}

@article{Guan2019OnLaw,
    title = {{On the AIC-based model reduction for the general Holzapfel–Ogden myocardial constitutive law}},
    year = {2019},
    journal = {Biomechanics and Modeling in Mechanobiology},
    author = {Guan, Debao and Ahmad, Faizan and Theobald, Peter and Soe, Shwe and Luo, Xiaoyu and Gao, Hao},
    number = {4},
    month = {8},
    pages = {1213--1232},
    volume = {18},
    publisher = {Springer Verlag},
    doi = {10.1007/s10237-019-01140-6},
    issn = {16177940},
    pmid = {30945052}
}

@article{Ahmed2022InterstitialRegurgitation,
    title = {{Interstitial Collagen Loss, Myocardial Remodeling, and Function in Primary Mitral Regurgitation}},
    year = {2022},
    journal = {JACC: Basic to Translational Science},
    author = {Ahmed, Mustafa I. and Andrikopoulou, Efstathia and Zheng, Jingyi and Ulasova, Elena and Pat, Betty and Kelley, Eric E. and Powell, Pamela Cox and Denney, Thomas S. and Lewis, Clifton and Davies, James E. and Darley-Usmar, Victor and Dell'Italia, Louis J.},
    number = {10},
    month = {10},
    pages = {973--981},
    volume = {7},
    publisher = {Elsevier Inc.},
    doi = {10.1016/j.jacbts.2022.04.014},
    issn = {2452302X}
}

@article{Spinale2007MyocardialFunction,
    title = {{Myocardial Matrix Remodeling and the Matrix Metalloproteinases: Influence on Cardiac Form and Function}},
    year = {2007},
    author = {Spinale, Francis G},
    journal = {Physiological reviews},
    volume = {87},
    issue = {4},
    pages = {1285–1342},
    doi = {10.1152/physrev.00012.2007}
}

@article{SahliCostabal2019MultiscaleFailure,
    title = {{Multiscale characterization of heart failure}},
    year = {2019},
    journal = {Acta Biomaterialia},
    author = {Sahli Costabal, F. and Choy, J. S. and Sack, K. L. and Guccione, J. M. and Kassab, G. S. and Kuhl, E.},
    month = {3},
    pages = {66--76},
    volume = {86},
    publisher = {Acta Materialia Inc},
    doi = {10.1016/j.actbio.2018.12.053},
    issn = {18787568},
    pmid = {30630123}
}

@article{Perry2002AngiotensinDog,
    title = {{Angiotensin II receptor blockade does not improve left ventricular function and remodeling in subacute mitral regurgitation in the dog}},
    year = {2002},
    journal = {Journal of the American College of Cardiology},
    author = {Perry, Gilbert J. and Wei, Chih Chang and Hankes, Gerald H. and Dillon, S. Ray and Rynders, Patricia and Mukherjee, Rupak and Spinale, Francis G. and Dell'Italia, Louis J.},
    number = {8},
    month = {4},
    pages = {1374--1379},
    volume = {39},
    doi = {10.1016/s0735-1097(02)01763-1},
    issn = {07351097},
    pmid = {11955858}
}

@article{Klotz2006DownloadedLib,
    title = {{Single-beat estimation of end-diastolic pressure-volume relationship:
a novel method with potential for noninvasive application}},
    year = {2006},
    journal = {Am J Physiol Heart Circ Physiol},
    author = {Klotz, Stefan and Hay, Ilan and Dickstein, Marc L and Yi, Geng-Hua and Wang, Jie and Maurer, Mathew S and Kass, David A and Burkhoff, Daniel},
    pages = {403--412},
    volume = {291},
    doi = {10.1152/ajpheart.01240.2005.-Whereas}
}

@Article{Rohmer2007,
  author   = {Rohmer, Damien and Sitek, Arkadiusz and Gullberg, Grant T.},
  journal  = {Investigative Radiology},
  title    = {Reconstruction and {Visualization} of {Fiber} and {Laminar} {Structure} in the {Normal} {Human} {Heart} from {Ex} {Vivo} {Diffusion} {Tensor} {Magnetic} {Resonance} {Imaging} ({DTMRI}) {Data}},
  year     = {2007},
  issn     = {0020-9996},
  month    = nov,
  number   = {11},
  pages    = {777},
  volume   = {42},
  doi      = {10.1097/RLI.0b013e3181238330},
  language = {en-US},
}

@article{Tanaka1986QuantitativeCardiomyopathy,
    title = {Quantitative analysis of myocardial fibrosis in normals, hypertensive hearts, and hypertrophic cardiomyopathy},
    year = {1986},
    journal = {British heart journal},
    author = {Tanaka, Masaru and Fujiwara, Hisayoshi and Onodera, Tomoya and Wu, Der-Jinn and Hamashima, Yoshihiro and Kawai, Chuichi},
    pages = {575--81},
    volume = {55},
    issue = {6},
    doi = {10.1136/hrt.55.6.575}
}

@article{Zhuan2019CoupledInfarction,
    title = {{Coupled agent-based and hyperelastic modelling of the left ventricle post-myocardial infarction}},
    year = {2019},
    journal = {International Journal for Numerical Methods in Biomedical Engineering},
    author = {Zhuan, Xin and Luo, Xiaoyu and Gao, Hao and Ogden, Ray W.},
    number = {1},
    month = {1},
    volume = {35},
    publisher = {Wiley-Blackwell},
    doi = {10.1002/cnm.3155},
    issn = {20407947},
    pmid = {30253447}
}

@article{Braeu2017HomogenizedRemodeling,
    title = {{Homogenized constrained mixture models for anisotropic volumetric growth and remodeling}},
    year = {2017},
    journal = {Biomechanics and Modeling in Mechanobiology},
    author = {Braeu, F. A. and Seitz, A. and Aydin, R. C. and Cyron, C. J.},
    number = {3},
    month = {6},
    pages = {889--906},
    volume = {16},
    publisher = {Springer Verlag},
    doi = {10.1007/s10237-016-0859-1},
    issn = {16177940},
    pmid = {27921189}
}

@article{Braeu2019AnisotropicTissues,
    title = {{Anisotropic stiffness and tensional homeostasis induce a natural anisotropy of volumetric growth and remodeling in soft biological tissues}},
    year = {2019},
    journal = {Biomechanics and Modeling in Mechanobiology},
    author = {Braeu, F. A. and Aydin, R. C. and Cyron, Christian J.},
    number = {2},
    month = {4},
    pages = {327--345},
    volume = {18},
    publisher = {Springer Verlag},
    doi = {10.1007/s10237-018-1084-x},
    issn = {16177940},
    pmid = {30413985}
}

@article{Kroon2009AAneurysms,
    title = {{A theoretical model for fibroblast-controlled growth of saccular cerebral aneurysms}},
    year = {2009},
    journal = {Journal of Theoretical Biology},
    author = {Kroon, Martin and Holzapfel, Gerhard A.},
    number = {1},
    month = {3},
    pages = {73--83},
    volume = {257},
    doi = {10.1016/j.jtbi.2008.10.021},
    issn = {00225193},
    pmid = {19027028}
}

@article{Wilson2013ParametricAneurysms,
    title = {{Parametric study of effects of collagen turnover on the natural history of abdominal aortic aneurysms}},
    year = {2013},
    journal = {Proceedings of the Royal Society A: Mathematical, Physical and Engineering Sciences},
    author = {Wilson, J. S. and Baek, S. and Humphrey, J. D.},
    number = {2150},
    month = {2},
    volume = {469},
    publisher = {Royal Society},
    doi = {10.1098/rspa.2012.0556},
    issn = {14712946}
}

@article{Baek2006AAneurysms,
    title = {{A theoretical model of enlarging intracranial fusiform aneurysms}},
    year = {2006},
    journal = {Journal of Biomechanical Engineering},
    author = {Baek, S. and Rajagopal, K. R. and Humphrey, J. D.},
    number = {1},
    month = {2},
    pages = {142--149},
    volume = {128},
    doi = {10.1115/1.2132374},
    issn = {01480731},
    pmid = {16532628}
}

@article{Zhong2006HypertrophicPathway,
    title = {{Hypertrophic growth in cardiac myocytes is mediated by Myc through a Cyclin D2-dependent pathway}},
    year = {2006},
    journal = {EMBO Journal},
    author = {Zhong, Weiguang and Mao, Songyan and Tobis, Scott and Angelis, Ekaterini and Jordan, Maria C. and Roos, Kenneth P. and Fishbein, Michael C. and De Albor{\'{a}}n, Ignacio Moreno and MacLellan, W. Robb},
    number = {16},
    month = {8},
    pages = {3869--3879},
    volume = {25},
    doi = {10.1038/sj.emboj.7601252},
    issn = {02614189},
    pmid = {16902412}
}

@article{Sack2018ConstructionDT-MRI,
    title = {{Construction and validation of subject-specific biventricular finite-element models of healthy and failing swine hearts from high-resolution DT-MRI}},
    year = {2018},
    journal = {Frontiers in Physiology},
    author = {Sack, Kevin L. and Aliotta, Eric and Ennis, Daniel B. and Choy, Jenny S. and Kassab, Ghassan S. and Guccione, Julius M. and Franz, Thomas},
    number = {MAY},
    month = {5},
    volume = {9},
    publisher = {Frontiers Media S.A.},
    doi = {10.3389/fphys.2018.00539},
    issn = {1664042X}
}

@article{Figueroa2009ASimulations,
    title = {{A Computational Framework for Fluid-Solid-Growth Modeling in Cardiovascular Simulations}},
    year = {2009},
    journal = {Comput Methods Appl Mech Eng},
    author = {Figueroa, C Alberto and Baek, Seungik and Taylor, Charles A and Humphrey, Jay D},
    doi = {10.1016/j.cma.2008.09.013}
}

@article{Sez2014,
   author = {Pablo Sáez and Estefania Peña and Miguel Angel Martínez and Ellen Kuhl},
   doi = {10.1007/s00466-013-0959-z},
   issn = {01787675},
   issue = {6},
   journal = {Computational Mechanics},
   pages = {1183-1196},
   publisher = {Springer Verlag},
   title = {Computational modeling of hypertensive growth in the human carotid artery},
   volume = {53},
   year = {2014}
}

@article{Klepach2012,
  title = {Growth and remodeling of the left ventricle: A case study of myocardial infarction and surgical ventricular restoration},
  volume = {42},
  ISSN = {0093-6413},
  url = {http://dx.doi.org/10.1016/j.mechrescom.2012.03.005},
  DOI = {10.1016/j.mechrescom.2012.03.005},
  journal = {Mechanics Research Communications},
  publisher = {Elsevier BV},
  author = {Klepach,  Doron and Lee,  Lik Chuan and Wenk,  Jonathan F. and Ratcliffe,  Mark B. and Zohdi,  Tarek I. and Navia,  Jose L. and Kassab,  Ghassan S. and Kuhl,  Ellen and Guccione,  Julius M.},
  year = {2012},
  pages = {134–141}
}

@article{Kuhl2004,
  author    = {Ellen Kuhl},
  title     = {Theory and Numerics of Open System Thermodynamics},
  url  = {https://nbn-resolving.de/urn:nbn:de:hbz:386-kluedo-17385},
  journal = {habilitation},
  institution = {Kaiserslautern - Fachbereich Maschinenbau und Verfahrenstechnik},
  type      = {habilitation},
  year        = {2004},
}

@article{Helfenstein2010OnMaterials,
    title = {{On non-physical response in models for fiber-reinforced hyperelastic materials}},
    year = {2010},
    journal = {International Journal of Solids and Structures},
    author = {Helfenstein, J. and Jabareen, M. and Mazza, E. and Govindjee, S.},
    number = {16},
    month = {8},
    pages = {2056--2061},
    volume = {47},
    doi = {10.1016/j.ijsolstr.2010.04.005},
    issn = {00207683}
}

@article{Gultekin2019OnMaterials,
    title = {{On the quasi-incompressible finite element analysis of anisotropic hyperelastic materials}},
    year = {2019},
    journal = {Computational Mechanics},
    author = {G{\"{u}}ltekin, Osman and Dal, Hüsnü and Holzapfel, Gerhard A.},
    number = {3},
    month = {3},
    pages = {443--453},
    volume = {63},
    publisher = {Springer Verlag},
    doi = {10.1007/s00466-018-1602-9},
    issn = {01787675}
}

@article{Sansour2008OnAnisotropy,
    title = {{On the physical assumptions underlying the volumetric-isochoric split and the case of anisotropy}},
    year = {2008},
    journal = {European Journal of Mechanics, A/Solids},
    author = {Sansour, Carlo},
    number = {1},
    month = {1},
    pages = {28--39},
    volume = {27},
    doi = {10.1016/j.euromechsol.2007.04.001},
    issn = {09977538}
}

\end{document}